\documentclass{aa}

\usepackage{booktabs}
\usepackage{threeparttable}
\usepackage{graphicx}
\usepackage{color}
\usepackage{amsmath}
\usepackage{epstopdf}
\usepackage{url}

\usepackage{xcolor}
\definecolor{green}{HTML}{33CC33}
\definecolor{red}{HTML}{FF3300}
\definecolor{blue}{HTML}{3333FF}

\usepackage[colorlinks=true,citecolor=blue,urlcolor=green]{hyperref} 

\usepackage{caption}
\usepackage{subcaption}
\usepackage{times}
\usepackage[english]{babel}
\usepackage{mathrsfs}

\renewcommand{\eqref}[1]{Eq.~\ref{#1}}
\newcommand{\fref}[1]{Fig.~\ref{#1}}
\newcommand{\sref}[1]{Sect.~\ref{#1}}

\newcommand{\ie}{i.\,e.} 
\newcommand{\ia}{i.\,a.} 
\newcommand{\eg}{e.\,g.}

\newcommand{\numax}{$\nu_{\rm max}$}
\newcommand{\kp}{\emph{Kepler}}

\renewcommand{\hat}{HAT-P-7}

\numberwithin{equation}{section}

\makeatletter
\def\maketag@@@#1{\hbox{\m@th\normalfont\normalsize#1}}
\makeatother

\usepackage{natbib,twoopt}
\usepackage[varg]{txfonts}
\bibpunct{(}{)}{;}{a}{}{,} 

\begin{document}

\title{Asteroseismic inference on the spin-orbit\\ misalignment and stellar parameters of HAT-P-7\vspace*{0.1cm}} 
\author{Mikkel~N.~Lund\inst{\ref{inst1}}\thanks{\email{mikkelnl@phys.au.dk}} 
\and Mia Lundkvist\inst{\ref{inst1},\ref{inst2}} 
\and Victor Silva Aguirre\inst{\ref{inst1}} 
\and \\ G\"{u}nter Houdek\inst{\ref{inst1}}
\and Luca Casagrande\inst{\ref{inst3}} 
\and Vincent Van Eylen\inst{\ref{inst1}} 
\and Tiago~L.~Campante\inst{\ref{inst5},\ref{inst1}}
\and \\ Christoffer Karoff\inst{\ref{inst4},\ref{inst1}} 
\and Hans~Kjeldsen\inst{\ref{inst1}} 
\and Simon~Albrecht\inst{\ref{inst1}}
\and William~J.~Chaplin\inst{\ref{inst5},\ref{inst1}}
\and \\ Martin~Bo~Nielsen\inst{\ref{inst6},\ref{inst7}} 
\and Pieter Degroote\inst{\ref{inst8}} 
\and Guy~R.~Davies\inst{\ref{inst5},\ref{inst1}} 
\and Rasmus~Handberg\inst{\ref{inst5},\ref{inst1}}
}

\institute{Stellar Astrophysics Centre, Department of Physics and Astronomy, Aarhus University, Ny Munkegade 120, DK-8000 Aarhus C, Denmark\label{inst1}
\and Sydney Institute for Astronomy (SIfA), School of Physics, University of Sydney, NSW 2006, Australia\label{inst2} 
\and Research School of Astronomy \& Astrophysics, Mount Stromlo Observatory, The Australian National University, ACT 2611, Australia\label{inst3}
\and Department of Geoscience, Aarhus University, H\o egh-Guldbergs Gade 2, 8000, Aarhus C, Denmark\label{inst4}
\and School of Physics and Astronomy, University of Birmingham, Edgbaston, Birmingham, B15 2TT, UK\label{inst5}
\and Institut für Astrophysik, Georg-August-Universität Göttingen, Friedrich-Hund-Platz 1, 37077 Göttingen, Germany\label{inst6} 
\and Max-Planck-Institut für Sonnensystemforschung, Justus-von-Liebig-Weg 3, 37077 Göttingen, Germany\label{inst7} 
\and Instituut voor Sterrenkunde, Katholieke Universiteit Leuven, Celestijnenlaan 200D, B-3001 Leuven, Belgium\label{inst8}}

\date{Received 03 June 2014 / Accepted 28 July 2014}


\abstract 
{The measurement of obliquities -- the angle between the orbital and stellar rotation -- in star-planet systems is of great importance for the understanding of planet system formation and evolution. The bright and well studied HAT-P-7 (Kepler-2) system is intriguing as several Rossiter-McLaughlin (RM) measurements found a large projected obliquity in this system, but it was so far not possible to determine if the orbit is polar and/or retrograde.} 
{The goal of this study is to measure the stellar inclination and hereby the full 3D obliquity of the HAT-P-7 system instead of only the 2D projection as measured by the RM effect. In addition we provide an updated set of stellar parameters for the star.} 
{We use the full set of available observations from \kp{} spanning Q$0$-Q$17$ to produce the power spectrum of HAT-P-7. We extract oscillation mode frequencies via an MCMC peak-bagging routine, and use the results from this to estimate the stellar inclination angle. Combining this with the projected obliquity from RM and the inclination of the orbital plane allows us to determine the stellar obliquity. Furthermore, we use asteroseismology to model the star from the extracted frequencies using two different approaches to the modelling where either the MESA or the GARSTEC stellar evolution codes are adopted.} 
{Using our updated asteroseismic modelling we find, \ia, the following stellar parameters for \hat{}: $M_{\star}=1.51^{+0.04}_{-0.05} \, M_{\sun}$,  $R_{\star}=2.00^{+0.01}_{-0.02} \, R_{\sun}$, and age $= 2.07^{+0.28}_{-0.23}$ Gyr. Our asteroseismic modelling offers a high precision on the stellar parameters, for instance is the uncertainty on age of the order ${\sim}11\%$. For the stellar inclination we estimate $i_{\star}<36.5^{\circ}$, which translates to an obliquity of $83^{\circ}<\psi<111^{\circ}$. We find that the planet HAT-P-7b is likely retrograde in its orbit, and that the orbit is close to being polar. The new parameters for the star gives an updated planetary density of $\rho_p=0.65\pm 0.03\, \rm g\, cm^{-3}$, which is lower than previous estimates.} {}


\maketitle

\titlerunning{HAT-P-7}
\authorrunning{Lund et al.}


\section{Introduction}
\label{sec:intro}

\begin{figure*}
        \begin{center}
        \begin{subfigure}[b]{0.4\textwidth}
                \includegraphics[trim=2.5cm 2.5cm 2.5cm 2.5cm, clip=true, width=\textwidth]{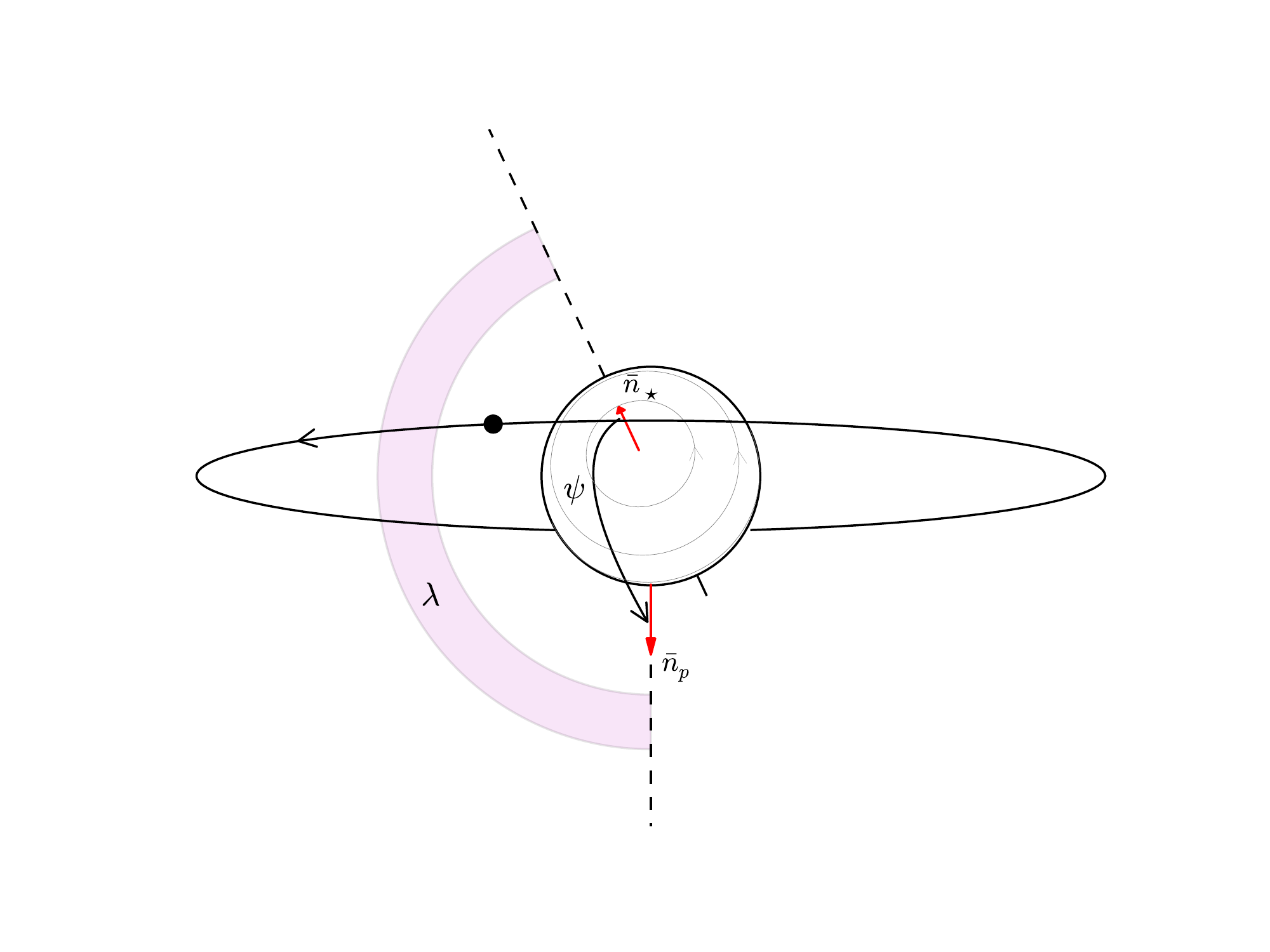}
                \caption*{Observers (front) view}
                \label{fig:system1}
        \end{subfigure} \quad
        \begin{subfigure}[b]{0.4\textwidth}
                \includegraphics[trim=2.5cm 2.5cm 2.5cm 2.5cm, clip=true, width=\textwidth]{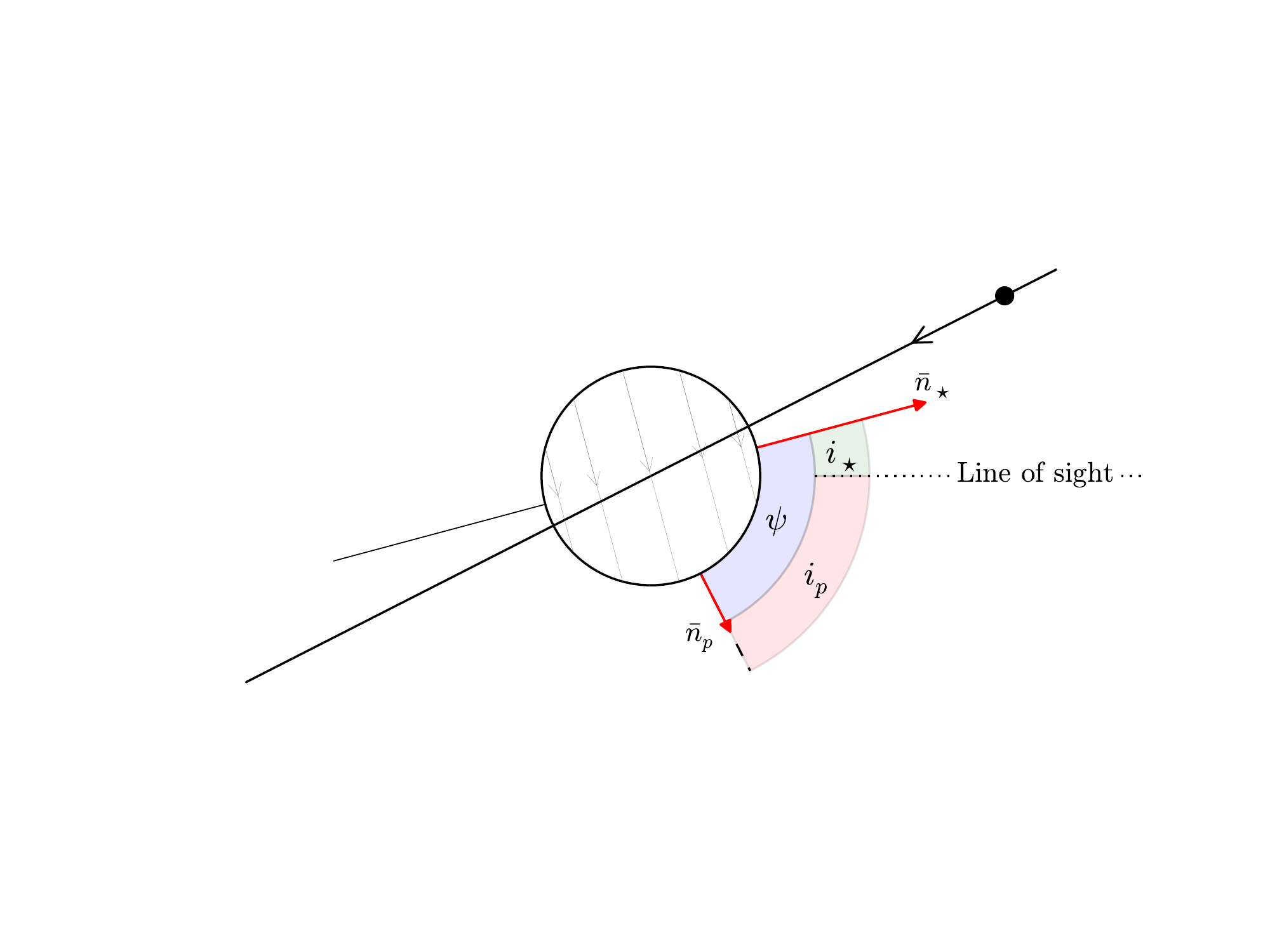}
                \caption*{Side view}
                \label{fig:system2}
        \end{subfigure}
        \end{center}
        \hspace{1.5cm}
        \begin{subfigure}[b]{0.4\textwidth}
                \includegraphics[trim=3.9cm 1.5cm 3.8cm 3.5cm, clip=true, width=\textwidth]{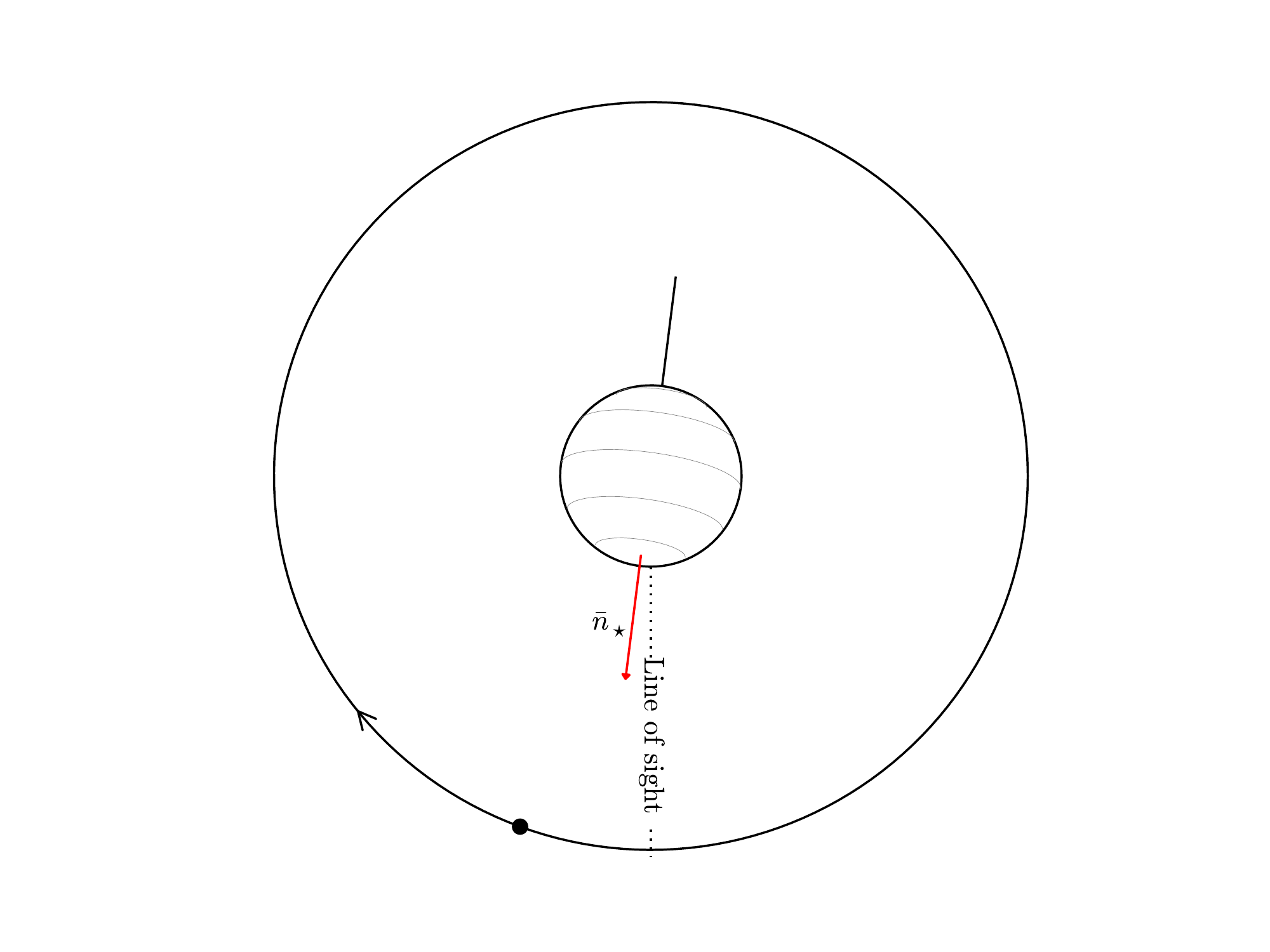}
                \caption*{Top view}
                \label{fig:system3}
        \end{subfigure}
\caption{\footnotesize Configuration of the HAT-P-7 system to scale using values for HAT-P-7b (full black circle) from \citet[][]{2013ApJ...774L..19V}. Top Left: the observers view of the system from Earth, with the angular momentum vectors of the planetary orbit, $\bar{n}_p$ (normal to the orbital plane), and stellar spin, $\bar{n}_{\star}$, given as red arrows. The projected angle, $\lambda$, is indicated by the shaded magenta region and is found as the angle between $\bar{n}_p$ and $\bar{n}_{\star}$ when these are projected onto the plane of the sky (dashed lines). This angle is obtained from RM measurements and in this panel we have used $\lambda=155\pm 37^{\circ}$ following \citet{2012ApJ...757...18A} (the uncertainty on $\lambda$ is not included in the figure). The stellar inclination, $i_{\star}$, which is the parameter measured from asteroseismology, is set to $15^{\circ}$, and is given by the direct angle between the line of sight (midpoint of star) and $\bar{n}_{\star}$. The inclination of the planetary orbit, $i_p$, is set to $83^{\circ}$ following \citet[][]{2013ApJ...774L..19V}. The true angle, $\psi$, is the direct angle between $\bar{n}_p$ and $\bar{n}_{\star}$. Top Right: side view of the system, with the observers view-point from the right, indicated by the ``line of sight''. To properly show $i_{\star}$ (shaded green) and $i_p$ (shaded red), and not their projected values, we have set $\lambda=180^{\circ}$ such that both $\bar{n}_p$ and $\bar{n}_{\star}$ lie in the same plane as the line of sight. For illustrative purposes we have in addition decreased $i_p$ to $63^{\circ}$. When adopting this configuration $\psi$ (shaded blue) is given by the sum of $i_{\star}$ and $i_p$.
Bottom Left: Top view of the system, with the observers view-point from the bottom, indicated by the ``line of sight''.  }     
\label{fig:system}
\end{figure*}

Asteroseismology can provide detailed information about stellar parameters such as mass, radius, and age \citep[see][and references therein]{2013ARA&A..51..353C}. Furthermore, an estimate for the stellar inclination can be obtained for solar-like oscillators \citep[][]{2003ApJ...589.1009G}, which in turn is needed in asserting the obliquity of planet hosting systems.
The obliquity of planetary systems, $\psi$, the angle between the stellar spin-axis and the angular momentum vector of the planetary orbit, is an important parameter for a better understanding of how these systems form and evolve \citep[see, \eg,][]{2008ApJ...678..498N,2010ApJ...718L.145W,2010A&A...524A..25T,2011ApJ...729..138M,2012ApJ...758L...6R}.
The obliquity is especially interesting for systems that appear to have retrograde orbits from measurements of the sky-projected obliquity, $\lambda$. The reason is that, while the orbit might indeed be retrograde, it is not known whether it is closer to polar than equatorial. This distinction makes a great difference for theories dealing with planetary system formation and evolution as they must be able to account for such a configuration \citep[see, \eg,][]{2008ApJ...686..580C,2008ApJ...686..621F,2010ApJ...725.1995M,2011MNRAS.412.2790L,2012ApJ...757...18A}. 

Stellar obliquities are difficult to measure as stars are unresolved by modern telescopes. Therefore no spacial information can be obtained\footnote{With the exception of stars with spacial interferometric constraints.}. However during planetary transits parts of the stellar surface are covered breaking the degeneracy. 
Such information on the obliquity can, \eg, be obtained from studies of the anomalous effect in the radial velocity (RV) curve known as the \emph{Rossiter-McLaughlin} (RM) effect \citep[][see \citet{2009ApJ...696.1230F} for an overview]{1924ApJ....60...15R, 1924ApJ....60...22M}.
Unfortunately, only $\lambda$ can be obtained from RM measurements (see \fref{fig:system}).
Other means of obtaining $\lambda$ are, \eg, from spot-crossing anomalies observed during planetary transits \citep[see, \eg,][]{2011ApJ...733..127S,2013ApJ...775...54S,2011ApJS..197...14D}, Doppler tomography \citep[see, \eg,][]{2012A&A...543L...5G}, or from the effects of gravity darkening \citep[see, \eg,][]{2011ApJS..197...10B,2014ApJ...786..131A}.

The true obliquity can only be unequivocally determined if the stellar angle of inclination, $i_{\star}$\footnote{We define $i_{\star}$ as the angle between the stellar spin axis and the observers line of sight, thus going from $i_{\star}=0^{\circ}$ for a pole-on view to $i_{\star}=90^{\circ}$ for an equator-on view.}, can be measured, and combined with $\lambda$ and the inclination of the planetary orbital plane, $i_p$.
The orbital inclination can be estimated with relative ease from analysis of the photometric light curve if the planet happens to transit its host star.

A measure of the stellar inclination angle can be obtained from combining $v\sin i_{\star}$ from spectroscopy with the stellar rotation period from modulations of the light curve from stellar spots \citep[see][for recent uses of this method for planetary systems]{2012ApJ...756...66H,2014ApJ...783....9H}. These estimates can be quite uncertain due to the difficulty of calibrating the spectroscopic $v\sin i_{\star}$, disentangling the rotational signal from other broadening effects, and the need for an estimate of the stellar radius $R_{\star}$. 
A more direct method for obtaining $i_\star$ is that of asteroseismology, where the stellar inclination can be estimated by analysing solar-like acoustic ($p$-mode) oscillations \citep[][]{2003ApJ...589.1009G}.
Another great advantage of using asteroseismology is that a detailed stellar model can be obtained with well determined parameters that are needed in simulations of planetary systems dynamics.   

The high photometric quality that enabled the \kp{} mission \citep[see][]{2010Sci...327..977B, 2010ApJ...713L..79K} to detect the transits of extrasolar planets, and thus allows for a determination of $i_p$, also make the data ideal for asteroseismic analysis \citep{2010PASP..122..131G}.
Analysis of the obliquity using the asteroseismic method have to date only been performed in the systems Kepler-50 and 65 by \citet[][]{2013ApJ...766..101C}, Kepler-56 by \citet[][]{2013Sci...342..331H}, Kepler-410 by \citet[][]{2014ApJ...782...14V}, and 16 Cygni by Davies et al. (submitted). However, for these systems the obliquity could only be assessed in a statistical sense, since the projected angle, $\lambda$, was unavailable.

Our aim in this paper is to use asteroseismology to determine precise stellar parameters of HAT-P-7 and to determine the obliquity of the HAT-P-7\footnote{We would like to emphasize the efforts made by O. Benomar and his collaborators for their work on HAT-P-7. This system turned out to be studied simultaneously and independently by our respective teams.} system with a F6V \citep[][]{2013MNRAS.433.2097F} type star and a close-in ${\sim}1.78\, M_{\rm J}$ planet (HAT-P-7b) in a ${\sim}2.2$ day orbit \citep{2008ApJ...680.1450P}. 
However, from the very onset it is clear that this is a challenging task as HAT-P-7 is a late-F-type star. This spectral type is notorious for having short lifetimes of the $p$-mode oscillations and consequently very broad (in frequency) oscillation modes, which highly obscures the potentially small effects imposed by rotation.

A fortuitous feature of the system is that not only $i_{\star}$ and $i_p$ can be estimated from \kp{} data, but the RM effect has been studied independently by \citet{2009ApJ...703L..99W}, \citet{2009PASJ...61L..35N}, and \citet[][]{2012ApJ...757...18A} using HIRES and/or HDS data\footnote{HIRES@Keck-I: High Resolution Spectrograph \citep[][]{1994SPIE.2198..362V};\\ HDS@Subaru: High Dispersion Spectrograph \citep[][]{2002PASJ...54..855N}.}. Values for $\lambda$ and $v\sin i_{\star}$ from these studies are given in Table~\ref{tab:orbit}. Despite disagreement on the actual values, all three works agree that the system is misaligned, that the planetary orbit might be retrograde, and that the very low value measured for $v\sin i_{\star}$ suggests a low $i_{\star}$, which for a transiting planet implies a near-polar orbit of the planet

An interesting aspect of the system with regard to the obliquity is that a third body (a M5.5V dwarf known as HAT-P-7B) is found to be associated with the system \citep[][]{2010PASJ...62..779N,2012PASJ...64L...7N,2013MNRAS.428..182B,2013MNRAS.433.2097F}, and that a fourth associated body, likely more massive than Jupiter, is speculated based on an unexplained RV excess \citep{2009ApJ...703L..99W,2012PASJ...64L...7N}.
The system is thus a prime candidate for obliquity studies and theories dealing with planetary system formation and evolution.

The paper is structured as follows: In \sref{sec:data} we describe the data used in our analysis. Section~\ref{sec:inc} deals with the model we use for obtaining the stellar inclination angle. In \sref{sec:res} we present our analysis and results, including stellar modelling using two different codes (\sref{sec:modelling}), and the results obtained for the stellar inclination and rotation (\sref{sec:splitinc}).
Our results for the obliquity is the topic of \sref{sec:obliquity}. In \sref{sec:gyro} we compare our result on the stellar rotation rate to gyrochronology, and \sref{sec:activity} is concerned with possible activity signatures. Finally we discuss our results in \sref{sec:dis} and make some concluding remarks in \sref{sec:con}.

\begin{figure*}
        \centering
        \begin{subfigure}[b]{0.45\textwidth}
                \includegraphics[width=\textwidth]{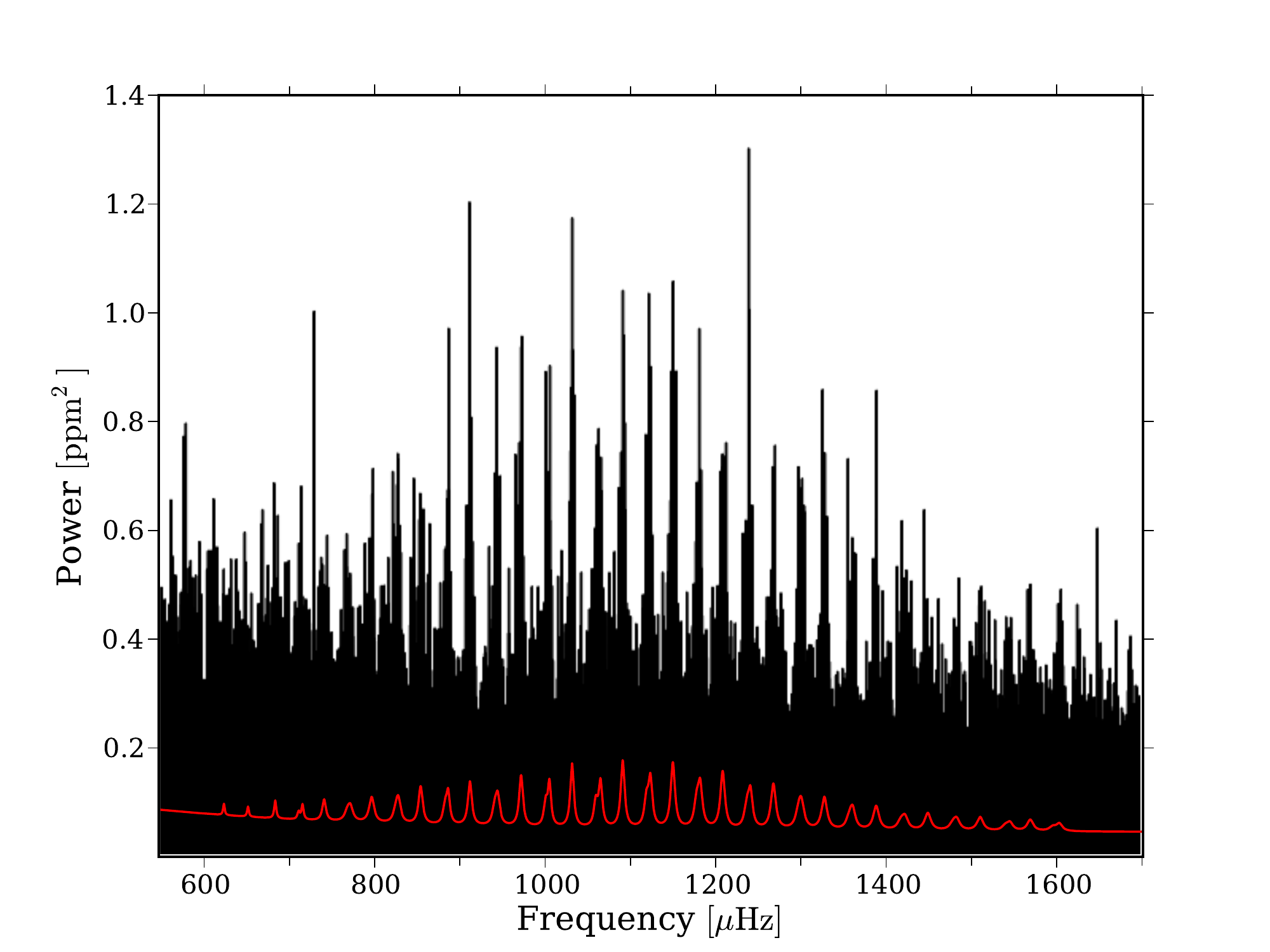}
                \label{fig:Power_spectrum}
        \end{subfigure} \quad
        \begin{subfigure}[b]{0.45\textwidth}
                \includegraphics[width=\textwidth]{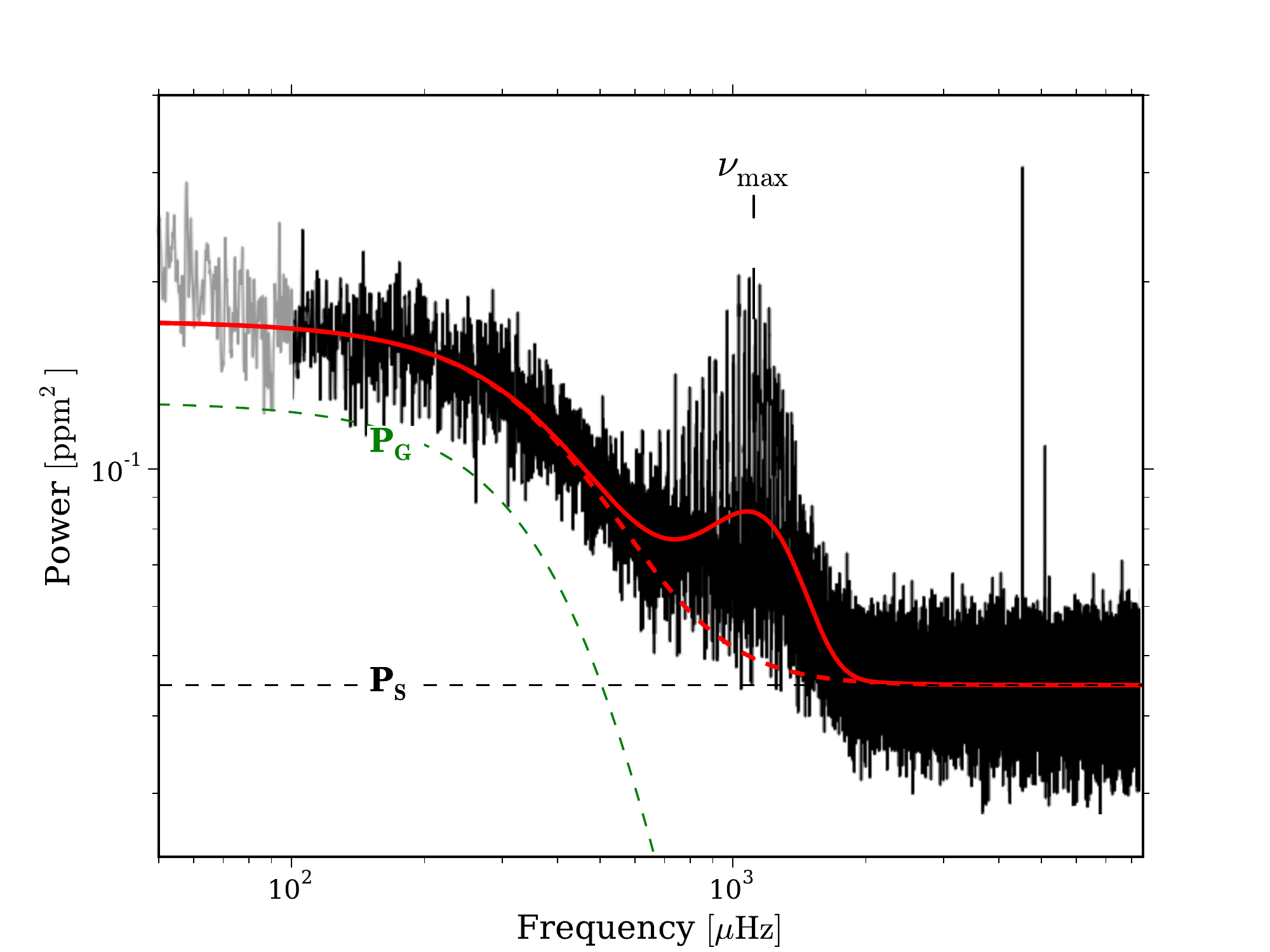}
                \label{fig:Background_fit}
        \end{subfigure}
\caption{\footnotesize Left: power spectrum of \hat{} (black) with the fitted model over-plotted (red; cf. \eqref{eq:limitspec}). Right: power spectrum of \hat{}\ (black) over-plotted with the optimum fit to the background (red; cf. \eqref{eq:bg}). The light-grey part up to $\rm 100\, \mu Hz$ was not included in the fit. The fit includes, besides the Gaussian envelope from $p$-modes centred around $\nu_{\rm max}\approx1115\,\rm \mu Hz$, a granulation component (\textcolor[rgb]{0, 0.5, 0}{\textbf{P}$\rm _G$}; green) and a white/shot noise (\textcolor{black}{\textbf{P$\rm _S$}}; black) level. The dashed red line shows the background fit without the Gaussian envelope.}     
\label{fig:Powerspecs_full}
\end{figure*}


\section{Data}
\label{sec:data}

We extracted short cadence (SC; $\Delta t = 58.8\,\rm s$) \emph{simple aperture photometry} (SAP) data from \emph{target pixel files} (TPFs) using the procedure of Steven Bloemen (private communication), which starts from the original \kp{} mask and adds or removes pixels to the aperture based on the amount of signal in each pixel. The outcome of this is a new mask that most often is slightly bigger than the original one. 

The data span quarters from Q$0$ to Q$17$ (${\sim}1470$ days), with a duty cycle of ${\sim}90.4\%$. This constitutes the full amount of data available from \kp{} in its normal mode of operation. 
Data were downloaded from \emph{The Mikulski Archive for Space Telescopes} (MAST) and corrected using the procedure described in Handberg \& Lund (submitted). Briefly, two median-filtered versions of the time series are computed with different filter windows. A weighted combination of the two are then used together with a filtered version of the planetary phase-curve to correct the time series for both instrumental features and the planetary signal.
For the asteroseismic analysis we used filter windows of $1$ ($\tau_{\rm long}$) and $0.028$ ($\tau_{\rm short}$) days, while windows of $15$ ($\tau_{\rm long}$) and $5$ ($\tau_{\rm short}$) days were adopted in preparing the time series used in \sref{sec:peak} to search for a low-frequency imprint of rotational modulation. We refer the reader to Handberg \& Lund (submitted) for further details on the filter and for a view of the corrected time series for HAT-P-7.

For the transit parameters needed in both the correction of the time series (orbital period) and for the estimation of the obliquity (inclination of orbital plane) we used the results of \citet[][]{2013ApJ...774L..19V}.
The power spectrum was calculated using a weighted sine-wave fitting method \citep[see, \eg,][]{1992PhDT.......208K,1995A&A...301..123F}, normalized according to the amplitude-scaled version of Parseval's theorem \citep[see][]{1992PASP..104..413K}, in which a sine wave of peak amplitude, A, will have a corresponding peak in the power spectrum of $\rm A^2$.


\section{Fitting the power spectrum}
\label{sec:inc}

The harmonic eigenmodes of acoustic solar-like oscillations are characterised by their degree, $l$, which gives the number of nodal lines on the stellar surface, and their radial order, $n$, giving the number of radial nodes. In addition, an eigenmode is characterised by its azimuthal order, $m$, of which there are $2l +1$. Only in the case of broken spherical symmetry, \eg\ by rotation, will the degeneracy between different $m$-values be lifted. The removal of this degeneracy makes it possible to measure the stellar rotation rate and inclination angle.  


\subsection{Modelling the power spectrum}
\label{sec:psmodel}

We model the power spectral density of the oscillations with a series of standard Lorentzian functions. The limit spectrum (noise free) to be fit to the power spectrum can be expressed as follows:
\begin{equation}
\mathcal{P}(\nu_j ; \boldsymbol \Theta) = \sum_{n=n_{a}}^{n_{b}}\sum_{l=0}^{2}\sum_{m=-l}^{l} \frac{\mathcal{E}_{l m}(i_{\star})S_{nl}}{1+\frac{4}{\Gamma_{nl}^2}(\nu_j-\nu_{nl m})^2 } + B(\nu_j)\, .
\label{eq:limitspec}
\end{equation}
Here $n_a$ and $n_b$ represent the first and last mode orders included from the power spectrum, respectively, while $\nu_{nl m}$ is the mode frequency including the effect of rotation, $B(\nu_j)$ describes the contribution from the stellar noise background at frequency $\nu_j$, $S_{nl}$ is the overall height of the multiplet, \ie, the maximum power spectral density, and $\Gamma_{nl}$ is the mode linewidth. A geometrical modulation of the relative visibility between components of a split multiplet is given by $\mathcal{E}_{l m}(i_{\star})$. The fitted parameters are denoted by $\boldsymbol \Theta$. This fit can be seen in the left panel of \fref{fig:Powerspecs_full}.

The background signal $B(\nu)$ is described by a series of power laws \citep{1985ESASP.235..199H}, each of which relate to a specific physical phenomenon. The power laws included in this work describe the signals from granulation and faculae. The specific functional from for the background signal is that suggested by \citet{KarPhD}:
\begin{align}
B(\nu) = \sum_{i=1}^2{\frac{4\sigma_i^2 \tau_i}{1 + (2\pi \nu \tau_i)^2 + (2\pi \nu \tau_i)^4}} + B_0 \, .
\label{eq:bg}
\end{align}
In this equation, $\sigma_i$ and $\tau_i$ gives, respectively, the flux rms variation in time and the characteristic time scales of the different phenomena. The constant $B_0$ is a measure of the photon shot-noise. The fit to the background can be seen in the right panel of \fref{fig:Powerspecs_full}.


\subsection{Optimisation procedure}
\label{sec:Optimisation_procedure}

The fit of \eqref{eq:limitspec} to the power spectrum is optimised in a Bayesian manner using the Markov Chain Monte Carlo (MCMC) sampler \texttt{emcee} (\citet[][]{2013PASP..125..306F}.

Frequencies for all modes are free parameters in the fit.
Heights and widths are only free parameters for radial ($l=0$) modes. For a non-radial mode the width is found from a linear interpolation between the two nearest radial modes. The same goes for the heights where the linearly interpolated value for the nearest radial modes is scaled via the visibility parameter (assumed constant as a function of frequency). In this work we keep the relative visibilities as free parameters.
With this set-up we have the following set of free parameters in the fitting: $\boldsymbol\Theta~=~\{\nu_{nl},\, i_{\star},\, \nu_s,\, S_{n,0},\, \tilde{V}_{l=1}^2,\, \tilde{V}_{l=2}^2,\, \Gamma_{n,0}\}$.

We refer the reader to Appendix~\ref{app:modelling_power} for details on the fitted model and the adopted optimisation.


\section{Analysis and Results}
\label{sec:res}


\subsection{Overall results from peak-bagging}
\label{sec:res_overall}

For estimating mode frequencies for the modelling we fit \eqref{eq:limitspec} to the frequency range $\rm 600 - 1650\, \mu Hz$, which is the range where we could visually identify modes.
All estimated frequencies are given in Table~\ref{tab:mcmc}. Here we report the median of the marginalized posterior (kernel) probability distributions (PPDs) for the respective modes, while the uncertainties were obtained from the $68\%$ \emph{highest probability density} (HPD) credible region. For the frequency uncertainties used in the stellar modelling (see \sref{sec:modelling}) we adopted the mean value of the (potentially asymmetric) uncertainties from the HPD credible region. 
For results on mode linewidths and visibilities we refer the reader to Appendix~\ref{app:lwvis}.

For the estimation of the inclination and splitting parameters we fit \eqref{eq:limitspec} to a smaller range in the power spectrum including only the frequencies in the range $\rm 780 - 1400\, \mu Hz$. The selection of this interval was based on the estimates for the mode linewidths from the large fit and to get modes of relatively high signal-to-noise ratio (SNR). For the splitting we used a flat prior from $\rm -8\, \text{to}\,8\, \mu Hz$, and then used the absolute value of the splitting in \eqref{eq:limitspec}. For the inclination we used a flat prior from $-90\, \text{to}\,180^{\circ}$, and then folded values onto the interval from $0\, \text{to}\,90^{\circ}$. The symmetry of the priors was chosen to avoid potential boundary effects in the MCMC sampling.

We note, that a component of \eqref{eq:limitspec} that can cause problems in the fitting is that of the noise-background. Initially we did not fix the background in the fit of \eqref{eq:limitspec}, but rather set Gaussian priors on the background parameters from the posteriors of a background-only fit. However, given that the background is very poorly constrained in the relatively small part of the power spectrum occupied by the oscillation modes, we found that not even the Gaussian priors could constrain the background, and especially the granulation time scale, $\tau_g$, wandered to lower values. These lower values for $\tau_g$ were found to correlate with a large range of high values for the splitting at a particular value for the inclination. Due to this apparent degeneracy we chose to fix the background in the fits.  


\subsection{Modelling of \hat{}}
\label{sec:modelling}

\begin{table*}
\begin{center}
\begin{threeparttable}
\caption{Selected properties of our best-fit models compared to the first asteroseismic modelling by \citet{2010ApJ...713L.164C} (CD10). The GARSTEC model (\sref{sec:GARSTEC}) constitutes our preferred values.}
\begin{tabular}{lllll}
\toprule \\ [-0.3cm] 
& & \multicolumn{2}{c}{This Work} \\
\cmidrule(r){3-4}
Parameter & & GARSTEC (preferred) &MESA &  CD10\tnote{a} \\[0.1cm]
\midrule
$M_\star\,\rm [M_\sun]$ 			&  & $1.51^{+0.04}_{-0.05}$		&$1.63 \pm 0.09$ 	&	 $1.52$			\\ [0.15cm]
$R_\star\,\rm [R_\sun]$ 			&  & $2.00^{+0.01}_{-0.02}$		&$2.04 \pm 0.04$ 	&	 $1.992$		\\ [0.15cm]
$L_\star\,\rm [L_\sun]$ 			&  & $5.91^{+0.31}_{-0.33}$		&$6.2 \pm 0.5$   	&	 $5.81$			\\ [0.15cm]
Age [Gyr]  						&  & $2.07^{+0.28}_{-0.23}$		&$1.9 \pm 0.4$   	&	 $1.875$		\\ [0.15cm]
$\log g\,\rm [cm\, s^{-2}]$ 		&  & $4.01^{+0.01}_{-0.01}$		&$4.03 \pm 0.01$ 	&	 $4.021$		\\ [0.15cm]
$T_\mathrm{eff} \,\rm [K]$ 		&  & $6366^{+78}_{-80}$			&$6374 \pm 130$  	&	  $6355$			\\ [0.15cm]
$[\mathrm{Fe}/\mathrm{H}]$ [dex]	&  & $+0.28^{+0.11}_{-0.11}$		&$+0.36 \pm 0.07$ 	&	 				\\ [0.15cm]
$Y_{\rm ini}$					&  & $0.288^{+0.009}_{-0.008}$ 	&$0.27 \pm 0.03$		&	 $0.2901$ 		\\ [0.15cm]
$X_{\rm ini}$ 					&  &	 $0.685^{+0.013}_{-0.015}$ 		&$0.70 \pm 0.03$		&  $0.6809$		\\ [0.15cm]
$M_{\rm core}\,\rm [M_\star]$\tnote{b} 	&  & $0.077^{+0.007}_{-0.008}$ 		&	 $0.076$			&  				\\ [0.15cm]
$\alpha$							&  & $1.791$ (fixed)				&$1.88 \pm 0.16$		&  $2.00$ (fixed)	\\ [0.15cm]
$f_\mathrm{ov}$					&  & $0.016$ (fixed)				&$0.003 \pm 0.002$	&  				\\ [0.15cm]
\bottomrule
\end{tabular}
\label{tab:mod_results}
\begin{tablenotes}
	\tiny
	\item [a] Values from the model with the smallest $\chi^2$, with convective overshoot over $0.1$ pressure scale heights included (model No.~2 in their Table~2). No uncertainties are reported by the authors.
	\item [b] Mass of the convective core from the position of the Schwarzschild boundary. The overshooting region extends beyond this point.
\end{tablenotes}
\end{threeparttable}
\end{center}
\end{table*}

\begin{figure}
\includegraphics[width=\columnwidth]{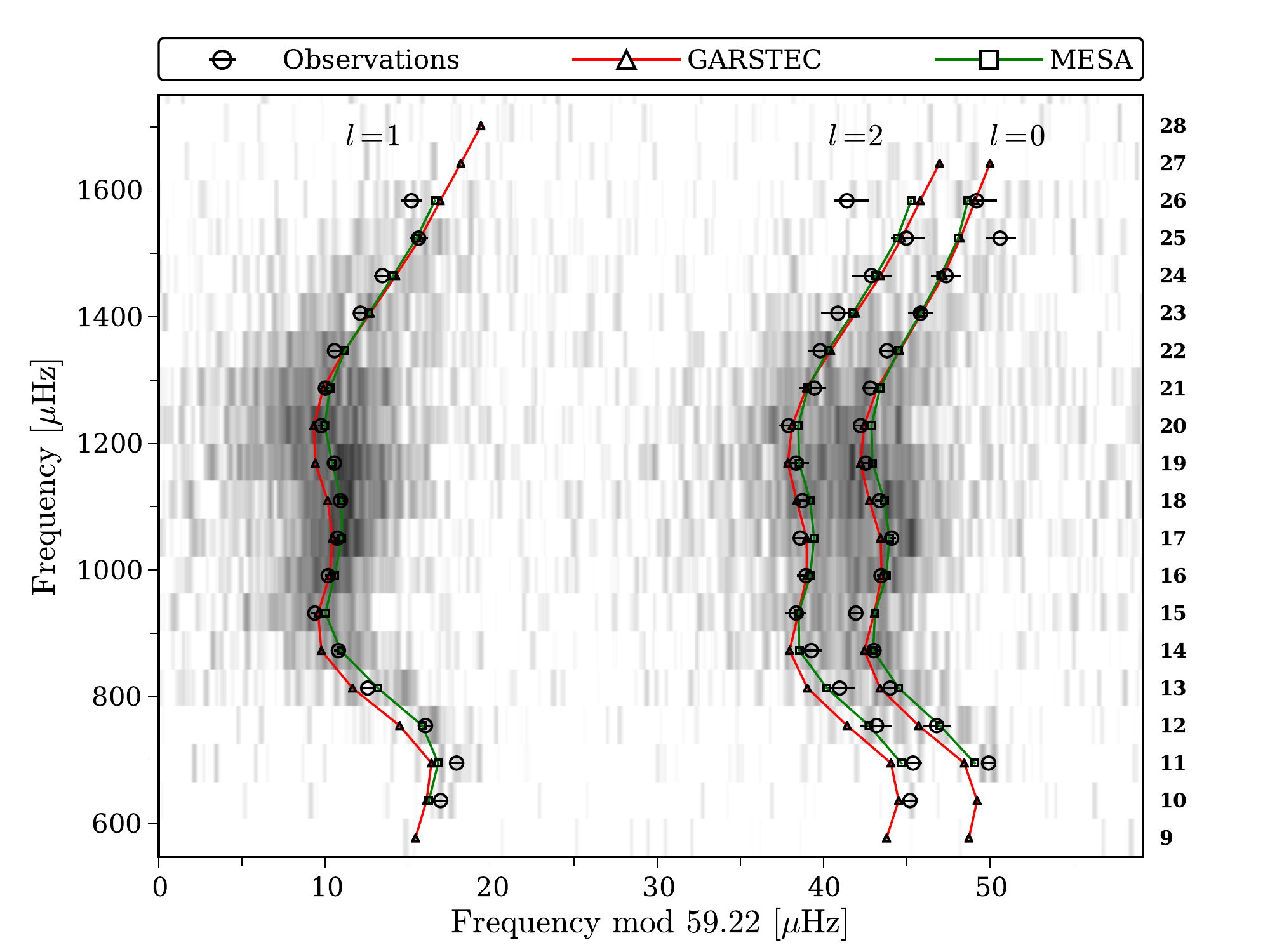}
\caption{\footnotesize \'{E}chelle diagram for \hat{} using $\rm \Delta\nu = 59.22\,\mu Hz$. The grey scale range from white at low power to black at high power. Circles give the extracted frequencies with corresponding uncertainties, triangles connected by red lines give model frequencies from GARSTEC, while squares connected by green lines give model frequencies from MESA. The degree of each ridge is indicated in the top part of the plot. The radial order of the $l=0$ modes is indicated by the numbers in the right side of the plot.}
\label{fig:model_compare}
\end{figure}

A detailed modelling of HAT-P-7 was first made by \citet{2010ApJ...713L.164C} based on asteroseismic measurements of the solar-like $p$-mode oscillations in the star. This work was based on the SC Q0-Q1 data from the \kp\ satellite. A fit of $33$ individual mode frequencies, with values obtained from peaks of a smoothed power spectrum, was made to models computed using ASTEC \citep{2008Ap&SS.316...13C}, with adiabatic pulsation frequencies calculated using ADIPLS \citep[Aarhus adiabatic oscillation package,][]{2008Ap&SS.316..113C}. 

Here we use the updated set of $50$ frequencies from our peak-bagging to model \hat{}, and use two different codes and modelling schemes to asses the robustness of our results. We note that this approach does not safeguard against other potential sources of systematics that can arise from, for instance, input physics not covered by the two codes or from assumptions made on certain quantities in the modelling. The effect of such systematics will be studied elsewhere (Silva~Aguirre et al. (in prep.)).

The results from the modelling can be found in Table~\ref{tab:mod_results}. The \'{e}chelle diagram \citep[][]{1983SoPh...82...55G} of \hat{} is given in \fref{fig:model_compare} after correcting for the background, and overlaid are both peak-bagged and modelled frequencies. In the construction of the \'{e}chelle diagram we have on the ordinate plotted the mid frequency of the respective $\Delta\nu$-length segments. Here $\Delta\nu$ denotes the so-called \emph{large-separation}, computed as the frequency difference between consecutive radial orders of a given degree. To obtain a better representation of the ridges for illustrative purposes a constant value was added to the frequencies before taking the modulo, thus allowing a shift of the ridges on the abscissa \citep[see, \eg,][]{2011arXiv1107.1723B}. The $\Delta\nu$ from \citet[][]{2013ApJ...767..127H} was used.


\subsubsection{MESA model}
\label{sec:MESA}

As a first approach \hat{} was modelled using MESA \citep[Modules for Experiments in Stellar Astrophysics,][]{ref:mesa11,ref:mesa13} and ADIPLS \citep[][]{2008Ap&SS.316..113C}. For the modelling we used the value of the effective temperature, $T_\mathrm{eff}\rm = 6350 \pm 126 \, K$, the heavy-element abundance, $[\mathrm{Fe}/\mathrm{H}] = +0.26 \pm 0.15$ \citep[both from][]{2014ApJS..211....2H}, and the frequencies given in Table~\ref{tab:mcmc}, except for the lowest $l = 0$ mode which gave consistently extremely poor agreement with the models and was subsequently excluded.

For the input physics we chose to neglect diffusion and settling following \citet{2010ApJ...713L.164C}. We used the 2005 update of the OPAL EOS \citep{ref:opal_1996,ref:opal_2002} and the NACRE nuclear reaction rates \citep{ref:nacre_reac_rates} with the updated $^{14}\mathrm{N}(p,\gamma)^{15}\mathrm{O}$ reaction rate by \citet{ref:14N} and the updated $^{12}\mathrm{C}(\alpha,\gamma)^{16}\mathrm{O}$ reaction rate by \citet{ref:12C}. Furthermore we used the OPAL opacities \citep{ref:opal_opacity} assuming the solar chemical composition given by \citet{ref:AGSS09}, supplemented by the \citet{ref:lowT_2005} opacities at low temperatures. We used the mixing-length theory of convection as formulated by \citet{ref:mixinglength}. Finally, we chose to use the "simple photosphere" option in MESA for the atmospheric boundary condition, which constitutes a grey atmosphere with the optical depth, $\tau_{\rm s}$, to the base of the atmosphere of $2/3$ \citep[see][their Equation~(3)]{ref:mesa11}.

In the matching of the model frequencies to the observed frequencies, we corrected the model frequencies for near-surface effects \citep{ref:surfaceterm} using the prescription by \citet{ref:brandao} with $b = 4.90$ and a reference frequency of $\rm \nu_0=1100 \, \mu Hz$. The best-fitting model was found via a $\chi^2$-minimisation \citep[see, \eg,][]{2010ApJ...713L.164C,ref:brandao,2013ApJ...763...49D}. The total $\chi^2$ was found as the weighted average of a normalised frequency and a spectroscopic component with weights given as $\chi^2 = \tfrac{2}{3}\chi_\nu^2 + \frac{1}{3}\chi_\mathrm{spectro}^2$ \citep[following][]{ref:mesa13} and the normalisation given by the number of values entering the $\chi^2$ component. The component $\chi_\mathrm{spectro}$ included the match of model and spectroscopic values for $T_\mathrm{eff}$ and $[\mathrm{Fe}/\mathrm{H}]$. The uncertainty of a given model parameter was found as the likelihood-weighted standard deviation of total $\chi^2$ values for all the computed models. Since we have chosen to use the model parameters for the best-fitting model instead of the likelihood-weighted mean values, we have added the difference between the mean and the best-fitting values in quadrature to the uncertainties.
The parameters of the best-fitting model and their uncertainties are listed in Table~\ref{tab:mod_results}.


\subsubsection{GARSTEC model}
\label{sec:GARSTEC}

\begin{figure}
\includegraphics[width=\columnwidth]{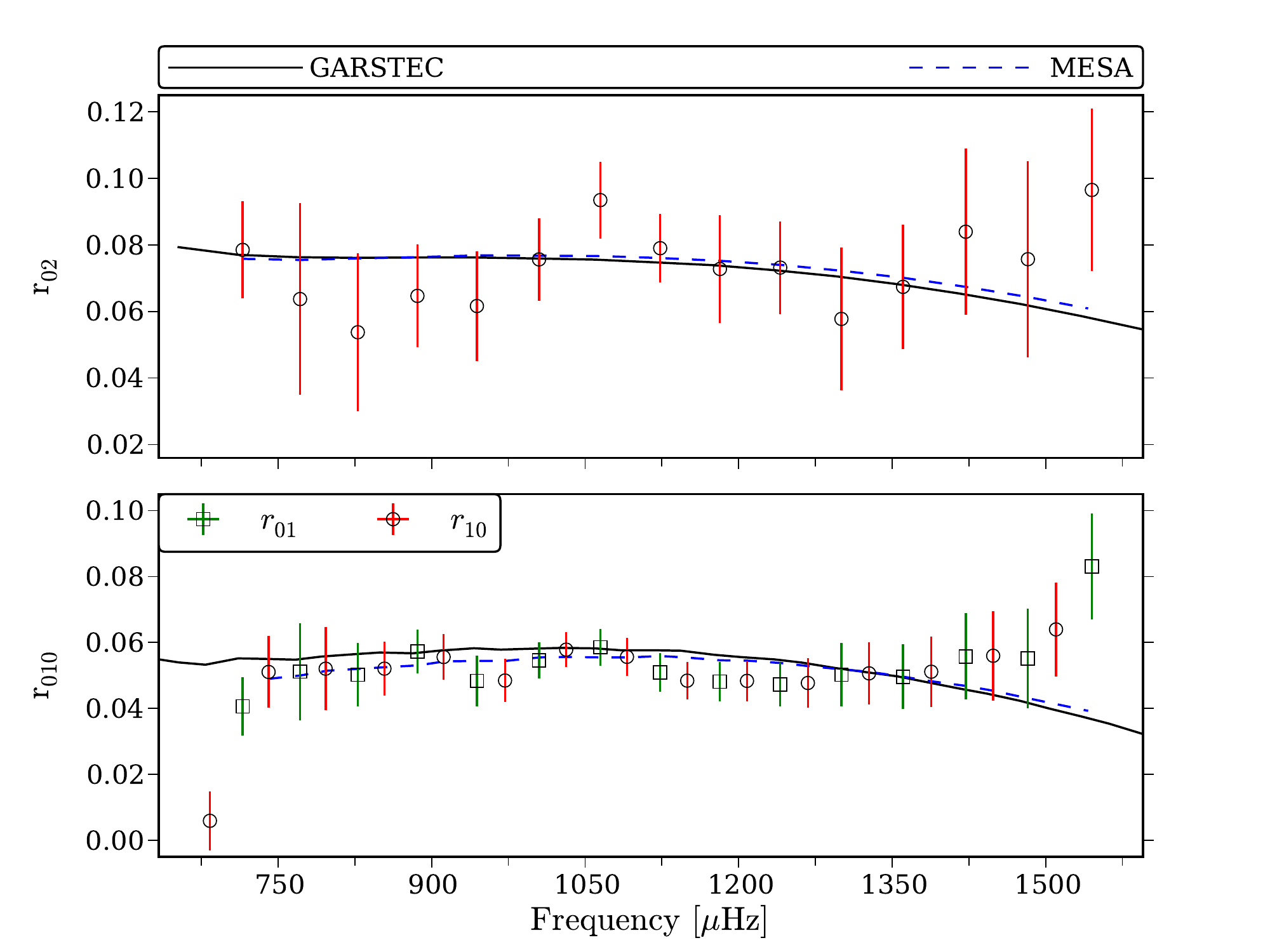}
\caption{\footnotesize Ratios $\rm r_{02}$ and $\rm r_{010}$ as a function of frequency. The lines show the ratios obtained for the best-fit GARSTEC (black solid) and MESA (dashed blue) models. Note that the relatively small uncertainties on frequencies (horizontal error bars) renders them undiscernible on the given scale.}
\label{fig:ratio}
\end{figure}

In the second approach we have used two grids of stellar models computed with the Garching Stellar Evolution Code \citep[GARSTEC, see][]{Weiss:2008jy}. 

The input physics is similar to the description given for the MESA model with the difference that we used the mixing-length theory of convection of \citet{2013sse..book.....K}, and \citet{Grevesse:1998cy} solar abundances. Also, the updated $^{12}\mathrm{C}(\alpha,\gamma)^{16}\mathrm{O}$ reaction rate was not used. 
In one grid we included the effects of core overshooting using an exponential decay of the convective velocities with an efficiency of $f_\mathrm{ov}=0.016$ \citep{Magic:2010iz}.

The grids cover a mass range between $0.7$ and $1.8\,M_{\sun}$ in steps of $0.01\,M_{\sun}$ and initial compositions of $-0.65<[\rm Fe/H]<+0.50$ in steps of 0.05~dex. These were determined using a galactic chemical evolution law of $\Delta Y/\Delta Z=1.4$ \citep[see, \eg,][]{2006AJ....132.2326B} anchored to the Big Bang nucleosynthesis value of $Y_p=0.248$ \citep{Steigman:2010gz}. For hundreds of models along each evolutionary track, we computed theoretical oscillations frequencies using ADIPLS \citep[][]{2008Ap&SS.316..113C}. These allowed us to construct dense grids of models covering the spectroscopic and asteroseismic parameter space of our target.

\begin{figure}
\centering
\includegraphics[scale=0.45, trim=0cm 1.2cm 0cm 0cm, clip=true,]{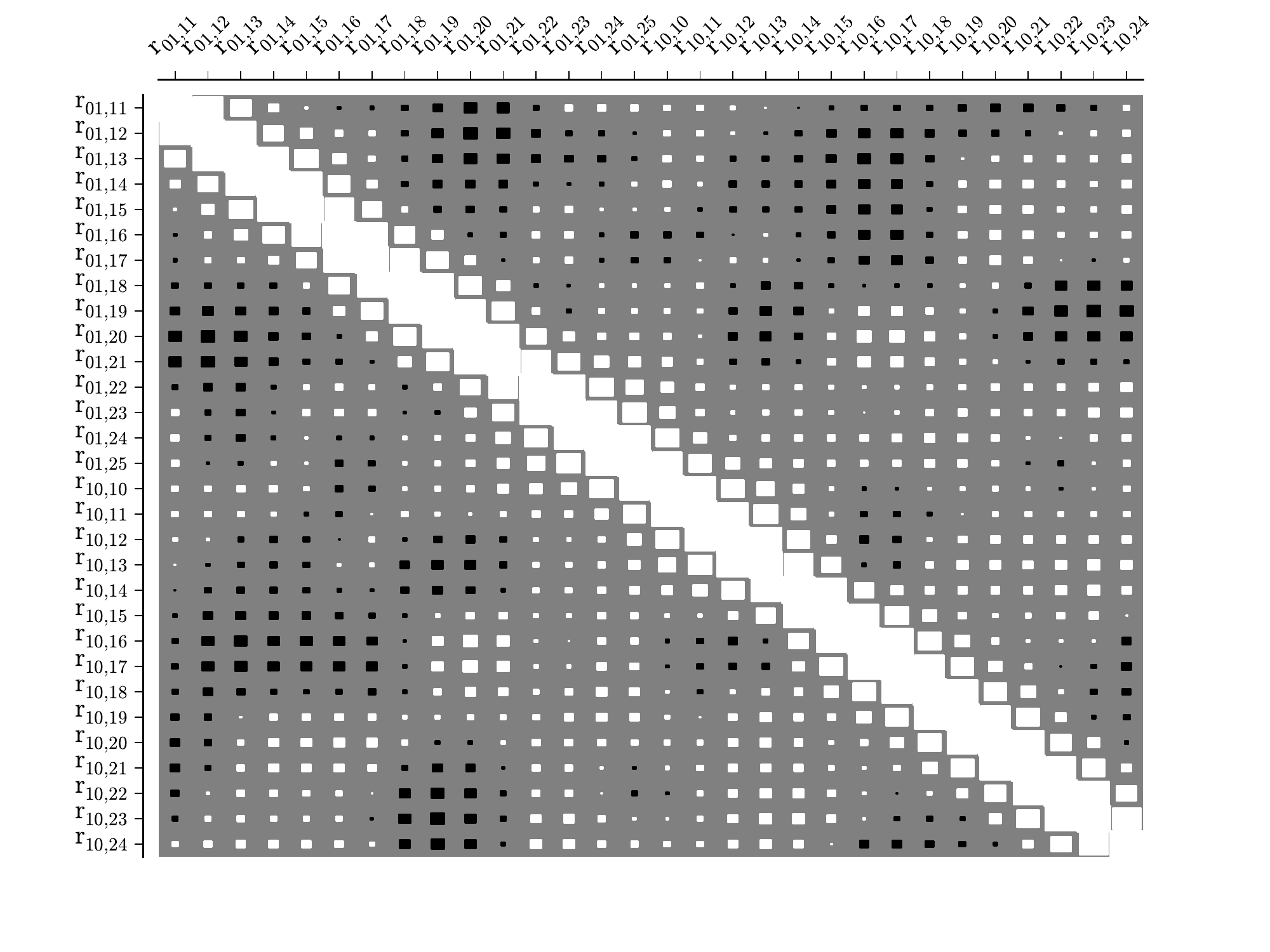}
\caption{\footnotesize Hinton diagram for the correlation matrix of $\rm r_{010}$ ratios. The first part of the subscript denotes if the ratio if of type $"01"$ or $"10"$, while the second gives the radial order of the central frequency. White (black) squares indicate positive (negative) covariances between ratios in question, and the size give the relative size of the correlation (one along the diagonal). }
\label{fig:hinton}%
\end{figure}

To determine the stellar parameters, we used the Bayesian approach described in Silva~Aguirre et al. (in prep.). Briefly, we assumed a flat prior in $\rm [Fe/H]$ and age including only a strict cut on the latter at $15\rm\, Gyr$, and we used a standard Salpeter IMF \citep[][]{1955ApJ...121..161S}. We computed the likelihood of the observed values given a set of model parameters assuming Gaussian distributed errors. In our case, the observables included were the spectroscopic $T_{\rm eff}$ and $\rm [Fe/H]$, and the frequency ratios defined as \citep[][]{2003A&A...411..215R}
\begin{align}
r_{01}(n) &= \frac{d_{01}(n)}{\Delta\nu_1(n)}, \quad r_{10}(n)=\frac{d_{10}(n)}{\Delta\nu_0(n+1)}\\ 
r_{02}(n) &=\frac{\nu_{n,0}-\nu_{n-1,2}}{\Delta\nu_1(n)}\, .
\end{align}
Here the $d_{01}$ and $d_{10}$ are the smooth 5-point small frequency separations given as
\begin{align}
d_{01}(n) &= \tfrac{1}{8} \left( \nu_{n-1,0} - 4\nu_{n-1,1} + 6\nu_{n,0} -4\nu_{n,1} +\nu_{n+1,0} \right)\\
d_{10}(n) &= -\tfrac{1}{8} \left( \nu_{n-1,1} - 4\nu_{n,0} + 6\nu_{n,1} -4\nu_{n+1,0} +\nu_{n+1,1} \right)\,.
\end{align} 
We refer to \citet[][]{SilvaAguirre:2013in} for further details.
In \fref{fig:ratio} we show the ratios obtained from the peak-bagging as a function of frequency along with ratios from the best-fit model. Here we also show the corresponding ratios obtained from the best-fit MESA model. Note that the ratios were not fitted in the MESA modelling. The construction of the ratios introduces correlations as a function of frequency that need to be taken into account when calculating the likelihood. 
This is done by calculating the $\chi^2$ entering the likelihood function as
\begin{equation}
\chi^2 = \frac{1}{N}\left(\boldsymbol{x}_{\rm obs} - \boldsymbol{x}_{\rm model} \right)^T \boldsymbol{\rm C}^{-1} \left(\boldsymbol{x}_{\rm obs} - \boldsymbol{x}_{\rm model} \right)\, ,
\end{equation}
where $\boldsymbol{\rm C}$ is the covariance matrix and the vectors $\boldsymbol{x}$ (of length $N$) give the observed and modelled values.
As an illustration of $\boldsymbol{\rm C}$ we show in \fref{fig:hinton} the Hinton diagram for the $\rm r_{010}$ ratios, which provides a qualitative view of the correlation matrix for these ratios. To construct ratios we use the PPDs from the individual frequencies. From these distributions we obtain our central values and uncertainties and use the Pearson standard correlation coefficient\footnote{Using the Python package \texttt{Pandas}. } to compute the correlation matrix. 

We found that the grid including core overshooting provided the best results.
The final set of parameters from the modelling are given in Table~\ref{tab:mod_results}. These were obtained from the median of the posterior probability distribution function and the 16 and 84 percent values. For the sake of comparison with the MESA model, the frequencies of the best-fit model are shown in \fref{fig:model_compare} after applying the \citet[][]{2008ApJ...683L.175K} method to correct for near-surface effects. Note, that the surface correction is not needed for this modelling approach as frequency ratios, not strongly affected by the surface layers \citep[][]{2003A&A...411..215R, 2011A&A...529A..63S}, are used rather than the actual frequencies. Consequently, any apparent misfit there might be between observed and surface corrected model frequencies in the \'{e}chelle diagram cannot simply be interpreted as being caused by a poor model, but could just as well be due to a poor surface correction. In the current case the fit, qualitatively speaking, reproduce the observations satisfactorily.

\subsubsection{Model comparison}
\label{sec:model_compare}

An improvement in our analysis compared to other works using \kp{} data is that now the full available dataset can be utilised. Furthermore, we find that the approach adopted in this work for the extraction of mode frequencies, \ie{} using an MCMC peak-bagging scheme, should provide more reliable estimates and uncertainties for the frequencies compared to the approach of assigning frequencies from peaks in a smoothed power spectrum.

In Table~\ref{tab:mod_results} we compare our results to the original asteroseismic results from \citet{2010ApJ...713L.164C}.
It is noteworthy that the GARSTEC model agrees quite well with these values, where only data from Q$0$-Q$1$ was used.
One contributing factor to the relatively small changes in model results from the addition of significantly more data is the F-type character of the star, where the frequency precision is limited by the large mode linewidths (see Appendix~\ref{app:lwvis}).  
Our MESA modelling results in a slightly more massive star than those obtained by GARSTEC and \citet{2010ApJ...713L.164C}, but still agree within uncertainties despite the different fitting techniques.  

The asteroseismic modelling by \citet{2012AN....333.1088V} resulted in a less massive ($M_{\star}=1.36\, M_{\sun}$) and less metal-rich ($[\mathrm{Fe}/\mathrm{H}]=0.13$) star than the current modelling efforts. This is likely because the authors explored a limited metallicity range and no spectroscopic constraints were included in the optimisation.
Our stellar parameters agree with those derived by \citet{2008ApJ...680.1450P} from a combined analysis of stellar isochrones, photometry, and spectroscopy. With our asteroseismic analysis we reduce the uncertainties on mass, radius, and age to $4.2\%$, $1.1\%$, and $11.2\%$ respectively from the original estimates of $6.4\%$, $13.9\%$, and $45.5\%$ given in \citet{2008ApJ...680.1450P}. In this comparison asymmetric uncertainties were added in quadrature.     

Since the GARSTEC values are computed using frequency ratios that are insensitive to near-surface effects, this set of stellar parameters will be adopted for the remainder of the paper. Furthermore, this Bayesian approach offers a direct estimation of the parameter uncertainties from the posterior distributions from a dense grid of models.

Finally, we note that the values of mass, radius, and age obtained from AME \citep{ref:ame}; $M/M_\sun = 1.55 \pm 0.05$, $R/R_\sun = 1.99 \pm 0.02$, $\mathrm{age} = 1.9 \pm 0.5 \, \rm Gyr$, are consistent with the results from this work.%


\subsubsection{Updated planetary parameters}
\label{sec:planet}

With our new estimates for the stellar parameters we can update the mass and radius for HAT-P-7b. The mass is found from \citep[see, \eg,][]{2010arXiv1001.2010W}
\begin{equation}
\frac{M_p}{(M_p + M_{\star})^{2/3}} = \frac{K_{\star}\sqrt{1-e^2}}{\sin i_p} \left( \frac{P_{\rm orb}}{2\pi G}\right)^{1/3}\, ,
\end{equation}
where $M_p$ is the planet mass, $e$ is the eccentricity, $P_{\rm orb}$ is the orbital period, and $K_{\star}$ is the stellar reflex velocity. As input we use $i_p$ and $P_{\rm orb}$ from \citet[][]{2013ApJ...774L..19V}, $e$ is set to zero following \citet[][]{2012MNRAS.422.3151H}, and $K_{\star}=212.2\pm 3.2\, \rm m\, s^{-1}$ from combining estimates of $K_{\star}$ by \citet{2009ApJ...703L..99W} and \citet{2009PASJ...61L..35N}. Solving for the planet mass then gives\footnote{Using $M_J = 1.899\,10^{30}\,\rm g$ and $R_J=7.1492\, 10^9\,\rm cm$.} $M_p=1.80\pm 0.05\, M_J$. For the planetary radius we get $R_p=1.51\pm 0.02\, R_J$ when using $R_p/R_{\star}$ from \citet[][]{2013ApJ...774L..19V}. Considering the discussion in \citet[][]{2013ApJ...774L..19V} on the size of systematic effects on transit depth measurements, we adopted a one per cent uncertainty on $R_p/R_{\star}$. This gives a planetary density of $\rho_p=0.65\pm 0.03\, \rm g\, cm^{-3}$, which is just within the uncertainty of the estimate from \citet{2008ApJ...680.1450P} of $\rho_p=0.876^{+0.17}_{-0.24}\, \rm g\, cm^{-3}$. We note that the stellar parameters estimated from asteroseismology offers a greatly improved precision on the planetary parameters.




\subsection{Splitting and inclination}
\label{sec:splitinc}

\begin{figure*}
\centering
\includegraphics[width=0.8\textwidth]{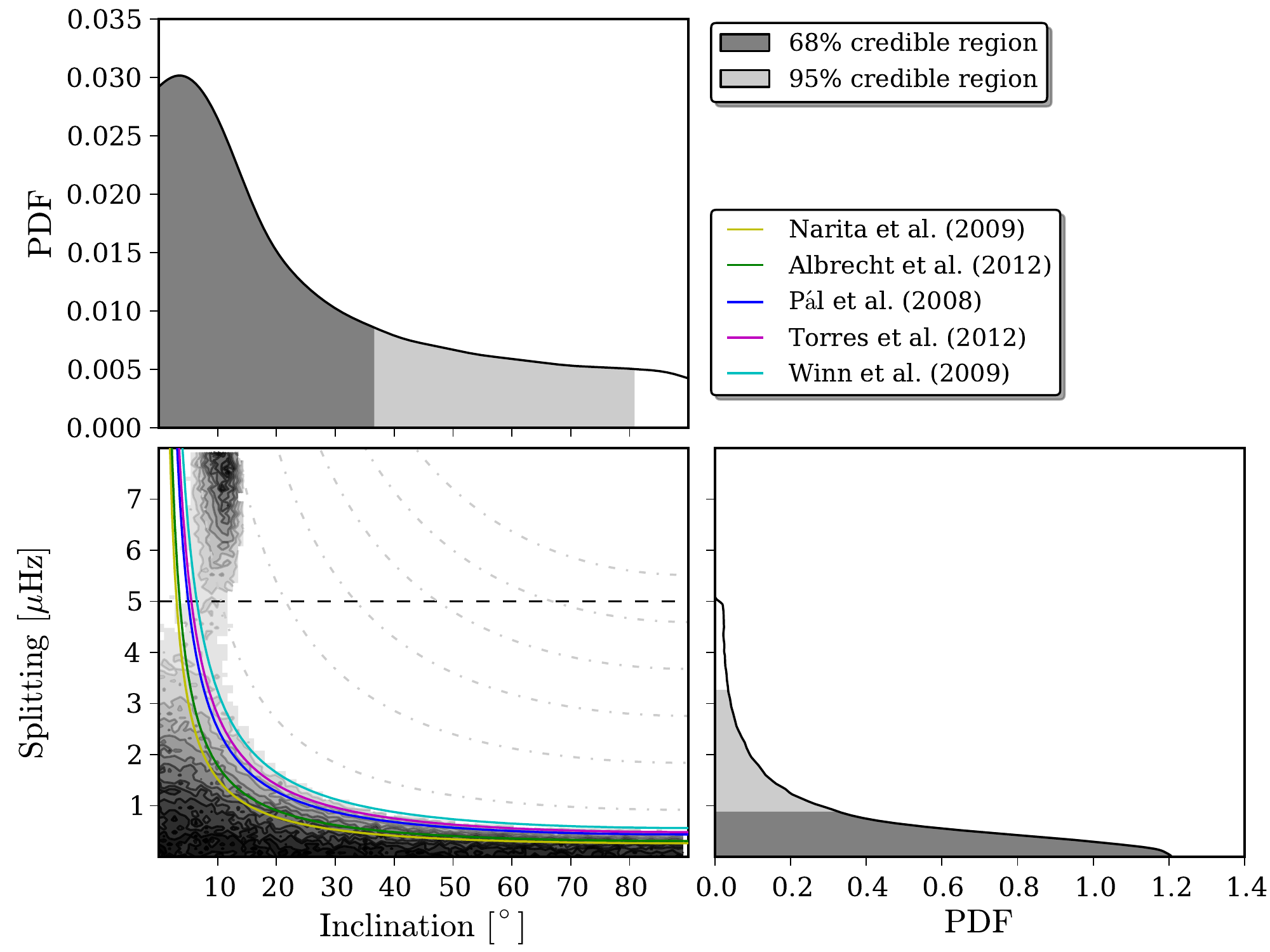}
\caption{\footnotesize Top: marginalized posterior probability distribution (PPD) for the stellar inclination $i_{\star}$. Here we have folded the full distribution ($-90$ to $180^{\circ}$) onto the the range from $0$ to $90^{\circ}$. Bottom right: PPD for the rotational splitting. Bottom left: correlation map between the inclination and the rotational splitting.
The $68\%$ credible regions (highest posterior density credible regions) are indicated by the dark gray part, while the light gray indicates the additional part covered in a $95\%$ credible region. Included are also lines of constant $v\sin i_{\star}$, computed using the radius estimate from our analysis and with $v\sin i_{\star}$ values from the literature (see Table~\ref{tab:orbit}), arranged in the legend in order of increasing $v\sin i_{\star}$ value (top to bottom). The dashed black horizontal line indicates the value of $\nu_s = 5\, \rm \mu Hz$ above which walkers were excluded from the PPDs. The grey dash-dot lines give lines of constant $v\sin i_{\star}$ from $8$ to $48\,\rm km\, s^{-1}$ in steps of $8\,\rm km\, s^{-1}$.}
\label{fig:bestfit}
\end{figure*}

In \fref{fig:bestfit} we present the results for the splitting and inclination of HAT-P-7 as the PPDs for these parameters, along with the 2D correlation map. These results were obtained from the small fit described above. Indicated in \fref{fig:bestfit} are the $68\%$ and $95\%$ HPD credible regions.

The distributions for inclination and splitting for HAT-P-7 are unfortunately not simple and Gaussian, but rather cover an extended region of parameter space. Considering this we will refrain from using measures of central tendency, such as the median or mode of the distributions, but rather report values from the $68\%$ HPD credible regions. The overall behaviour seen in most of the correlation map correspond well to what might be expected given the relatively large linewidths: $\rm\Gamma = 4.8 - 8.4\, \mu Hz$ for the range fitted (see \fref{fig:linewidth}). At low inclinations, the central $m=0$ component dominates the relative heights of azimuthal components of a rotationally split multiplet, while at high inclinations the sectoral ($m=\pm l$) components dominate. For $l=2$ the tesseral ($0<|m|<l$) components dominate at an inclination of ${\sim}50^{\circ}$. Thereby, at low inclinations the splitting is generally less constrained as the $m\neq0$ components are small and harder to fit. Furthermore, a linewidth exceeding the splitting effectively hides these small components up to a splitting of half the linewidth (\ie, up to $\rm{\sim}4\, \mu Hz$). At higher inclinations the splitting is more easily discernible as the visible azimuthal components are the ones furthest apart, thus the linewidth becomes less of a problem.

The reader will notice a high density at $i_{\star} \sim 10^{\circ}$ and $\nu_s > 5\, \rm \mu Hz$. This should not be interpreted as a good solution to the fit of the power spectrum but is rather the result of walkers trying to escape the parameter space, with the consequence that the walkers pile up at the boundary of the prior on the splitting. A contributing factor could also be that the small separation between $l=0$ and $l=2$ modes is of the order ${\sim}5\, \rm \mu Hz$, whereby $l=2$ azimuthal components would be within the $l=0$ mode profile. We see no correlations between these high splitting values and any of the other parameters of the fit and so these parameters are unaffected by this feature. These walkers will, however, contribute to the shape of the splitting and inclination PPDs and therefore we chose to exclude walkers with $\nu_s > 5\, \rm \mu Hz$ in the marginalization of the $i_{\star}$ and $\nu_s$ PPDs.
This is the same effect we observed when the background was kept free in the fit, but now much decreased. It is, however, not clear to us what is special about the inclination of $i_{\star} \sim 10^{\circ}$ in facilitating this escape of walkers.

From the $68\%$ HPD credible regions of the inclination and splitting PPDs we find $i_{\star}<36.5^{\circ}$ and $\nu_s < 0.87 \, \rm \mu Hz$ (see Table~\ref{tab:astro}). The value for the inclination is in overall agreement with the notions made by others based on the low value for $v\sin i_{\star}$. In addition, agreement is found with the statistically derived estimate by \citet[][]{2010ApJ...719..602S} of $i_{\star} = 6.7 - 10^{\circ}$ based on a simple $P_{\rm rot}\propto t^{1/2}$ model \citep[][]{1967ApJ...148..217W,1972ApJ...171..565S} for the evolution of the stellar rotation period as a function of mass and age.
The splitting, on the other hand, is lower than expected for an F6-type star (see below).
However, it is clear from the correlation map that at low inclinations, values of the splitting up to $\rm{\sim}4\, \mu Hz$ is allowed.

\begin{table}
\begin{center}
\begin{threeparttable}
\caption{Values related to the stellar inclination and rotation. All values are estimated from the $68\%$ HPD credible region of their corresponding parameter distributions.}
\begin{tabular}{ll}
\toprule \\ [-0.3cm] 
Parameter & $68\%$ HPD limit \\[0.1cm]
\midrule
$i_{\star}\, [^{\circ}]$ & \hspace{0.5cm}$<36.5$ \\ [0.2cm]
$\rm \nu_s\, [\mu Hz]$  & \hspace{0.5cm}$<0.87$ \\ [0.2cm]
$P_{\rm rot}\,\rm  [days]$  & \hspace{0.5cm}$>13.23$ \\ [0.2cm]
$v\sin i_{\star}\, [\rm km\, s^{-1}]$ & \hspace{0.5cm}$< 2.21$ \\ [0.2cm]
$v_{\rm surf}\, [\rm km\, s^{-1}]$ & \hspace{0.5cm}$<7.66$ \\ [0.2cm]
\bottomrule
\end{tabular}
\label{tab:astro}
\end{threeparttable}
\end{center}
\end{table}

\begin{figure}[h]
\includegraphics[width=\columnwidth]{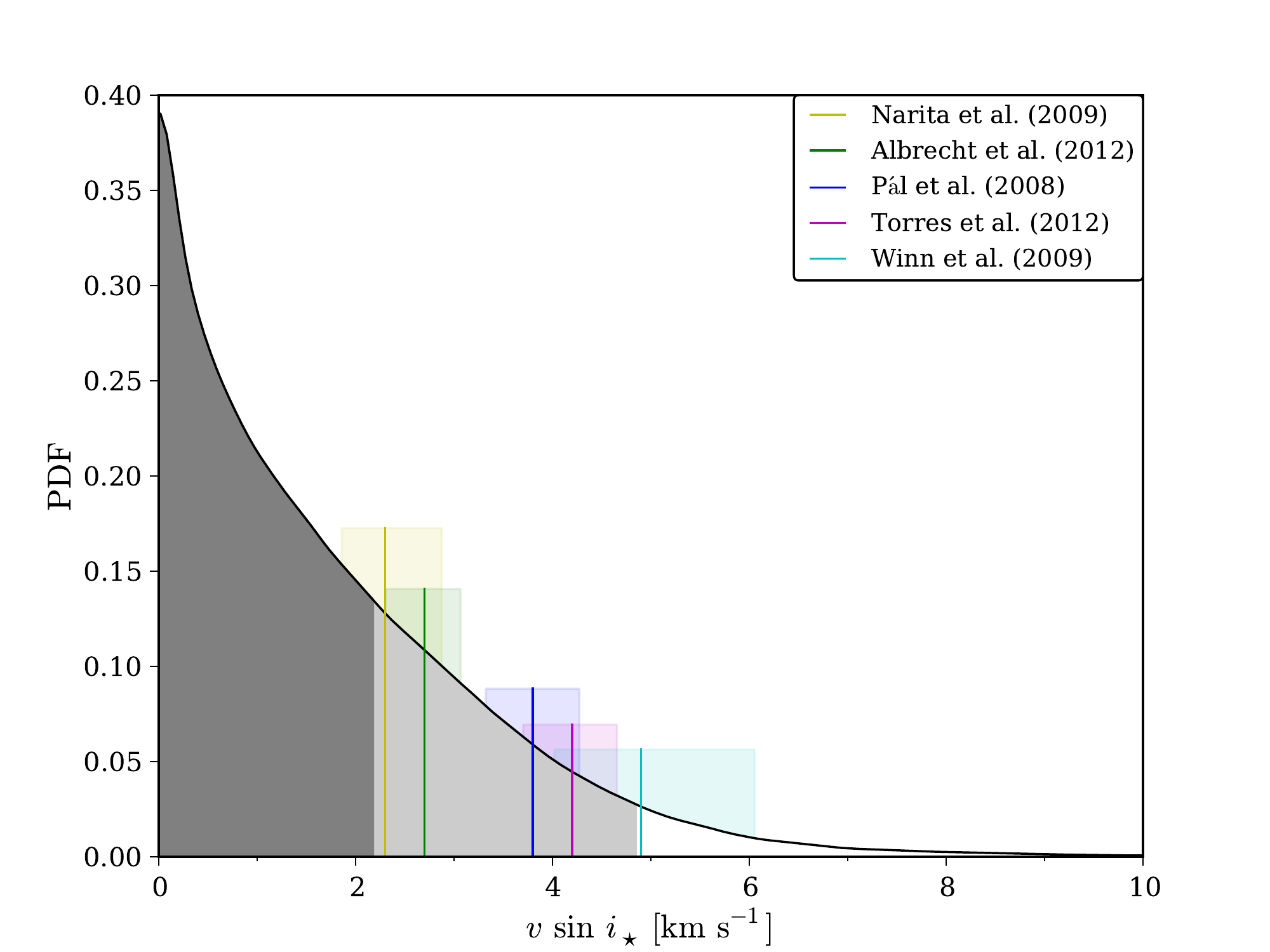}
\caption{\footnotesize Distribution for $ v\sin i_{\star}$ constructed from the PPDs obtained for $\nu_s$ and $i_{\star}$ and assuming $R/R_{\odot}\sim \mathcal{N}(2.00, 0.02)$ (see \eqref{eq:vsini}). The dark and light grey regions of the distribution corresponds to the $68\%$ and $95\%$ credible regions, respectively. Literature values for $v\sin i_{\star}$ obtained from spectral and RM analysis (see Table~\ref{tab:orbit}) are given by vertical lines where the shaded regions above the distribution give the corresponding uncertainties.}
\label{fig:vsini}
\end{figure}

With our estimate for the stellar radius (see \sref{sec:modelling}) we may convert the estimated ranges for the splitting and inclination to a measure of $v\sin i_{\star}$ as \citep[][]{2013ApJ...766..101C}
\begin{equation}
v\sin i_{\star} = 2\pi R \nu_s \sin i_{\star}\, .
\label{eq:vsini}
\end{equation}
Using the PPDs obtained for $\nu_s$ and $i_{\star}$, while assuming a distribution for the radius as $R/R_{\odot}\sim \mathcal{N}(2.00, 0.02)$, we find value of  $ v\sin i_{\star} < 2.21$ km s$^{-1}$ ($68\%$ HPD credible region). In \fref{fig:vsini} we have plotted the full distribution for $ v\sin i_{\star}$ and indicated the values obtained from spectral and RM analysis (see Table~\ref{tab:orbit}). We find that the value of \citet{2009PASJ...61L..35N} agree within uncertainties with our results. \citet{2009ApJ...703L..99W} found the largest value for $ v\sin i_{\star}$. Their higher $ v\sin i_{\star}$ value results from the lower orbital inclination value (see Table~\ref{tab:orbit}) they used for modelling the RM effect. For HAT-P-7 a larger impact parameter (lower inclination) requires a larger $ v\sin i_{\star}$. If one would repeat the analysis with the inclination derived from the \kp{} light curve, instead of the ground
based data available to them, then this disagreement in $ v\sin i_{\star}$ would vanish.

Having $ v\sin i_{\star}$ and $i_{\star}$ measured and assuming solid body rotation we can now calculate the true rotation speed of HAT-P-7, $v<7.66 \,\rm km\, s^{-1}$ ($68\%$ HPD credible region).
This upper limit estimate agrees with the values from \citet[][]{2013A&A...557L..10N} from the rotation periods of \kp{} stars of an approximate spectral type between F4 and 6.

\section{Obliquity}
\label{sec:obliquity}

\begin{table*}
\begin{center}
\begin{threeparttable}
\caption{Literature Values for Parameters Related to the \hat{} System.}
\begin{tabular}{lcccc}
\toprule \\ [-0.3cm] 
Source & $v\sin i_{\star}$ [km s$^{-1}$] & $\lambda\, [^{\circ}]$ & $i_{p}\, [^{\circ}]$  & $\psi$ (this work) \\[0.1cm]
\midrule
\citet{2008ApJ...680.1450P} 			& 3.8 $\pm$ 0.5  		&    						& 85.7$^{+3.5}_{-3.1}$	 & \\ [0.2cm]
\citet{2009ApJ...703L..99W}			& 4.9$^{+1.2}_{-0.9}$	& 182.5 $\pm$ 9.4			& 80.8$^{+2.8}_{-1.2}$	 & $83^{\circ}<\psi<119^{\circ}$\\[0.2cm]
\citet{2009PASJ...61L..35N}			& 2.3$^{+0.6}_{-0.5}$  	& -132.6$^{+10.5}_{-16.3}$\tnote{a} 	& 85.7$^{+3.5}_{-3.1}$	 & $83^{\circ}<\psi<106^{\circ}$\tnote{b}\\[0.2cm]
\citet{2012ApJ...757...18A}			& 2.7 $\pm$ 0.4  		& 155 $\pm$ 37\tnote{c}		& 					 	 & $83^{\circ}<\psi<111^{\circ}$\\[0.2cm]
\citet{2012ApJ...757..161T}			& 4.2$\pm$ 0.5 			& 							&   	 					 & \\[0.2cm]
\citet{2013ApJ...774L..19V}			&   					& 						& 83.151$^{+0.030}_{-0.033}$  	 & \\[0.2cm]
This work							& $< 2.21$ 			& 							&   	 					 & \\[0.1cm]
\bottomrule
\end{tabular}
\label{tab:orbit}
\begin{tablenotes}
	\tiny
	\item [a] The value for the projected obliquity is equivalent to $\lambda=227.4^{+10.5\circ}_{-16.3}$.
	\item [b] We used an uncertainty on $\lambda$ of $\pm 16.3^{\circ}$.
	\item [c] From a fit to RM data \citet{2012ApJ...757...18A} obtained an uncertainty on $\lambda$ of $\pm14^{\circ}$. The uncertainty adopted here was ascribed by \citet{2012ApJ...757...18A} as the standard deviation of the three independent measurements of $\lambda$. 
\end{tablenotes}
\end{threeparttable}
\end{center}
\end{table*}

\begin{figure}[h]
\includegraphics[width=\columnwidth]{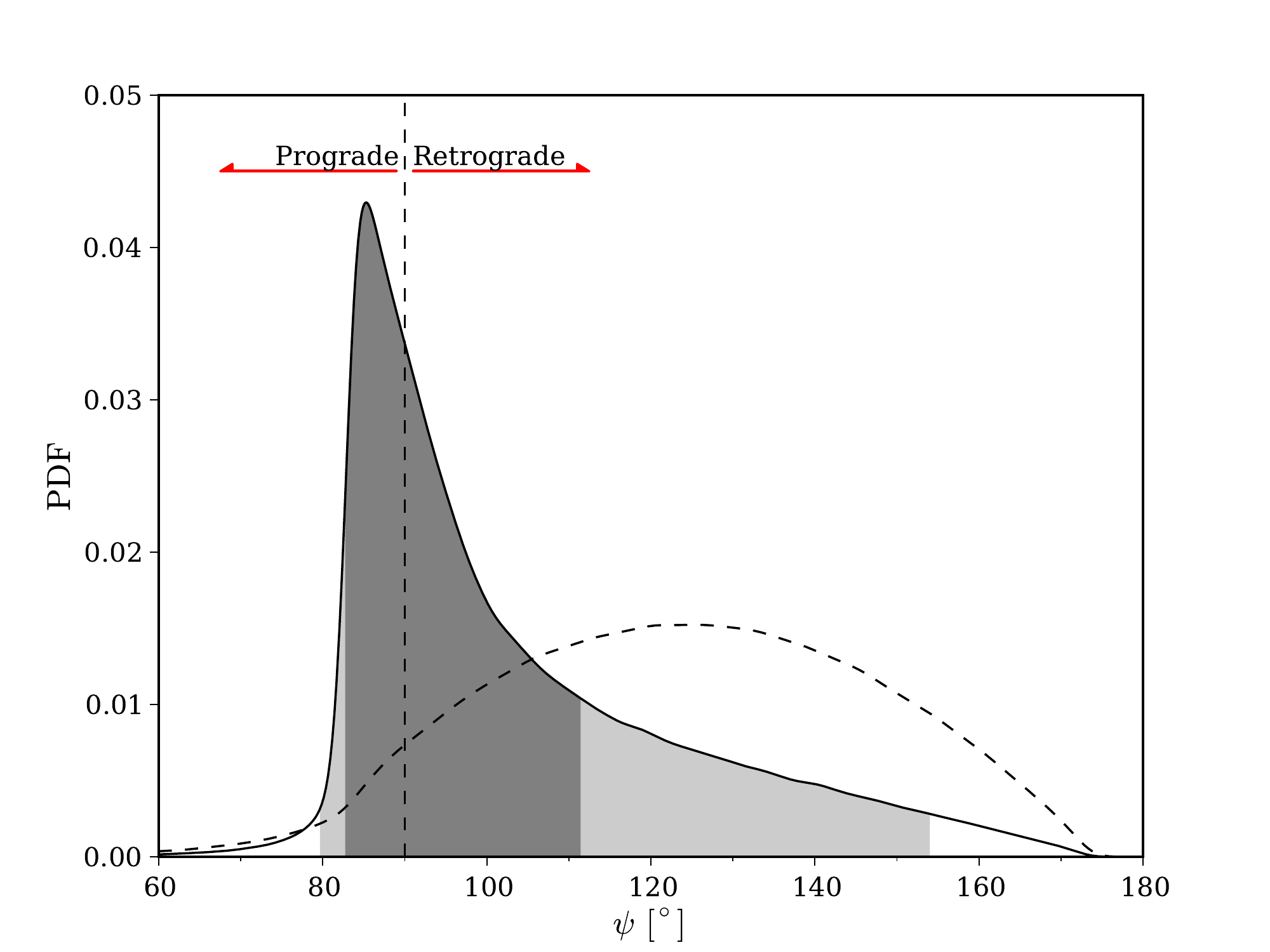}
\caption{\footnotesize Distribution of the true angle $\psi$ using the distribution for $i_{\star}$ from the peak-bagging, and with the assumption of Normal distributions for the planetary inclination $i_p$ and projected angle $\lambda$. For $i_p$ the value from \citet{2013ApJ...774L..19V} was adopted, while $\lambda$ was taken from \citet{2012ApJ...757...18A}. The dark and light grey regions of the distribution corresponds to the $68\%$ and $95\%$ credible regions, respectively. The dashed curve gives the distribution if one instead assumes an isotropic distribution for $i_{\star}$, \ie, flat in $\cos i_{\star}$. Indicated is also which values of $\psi$ that corresponds to a retrograde or prograde orbit of HAT-P-7b. }
\label{fig:ob}
\end{figure}

With our estimate for the stellar inclination, $i_{\star}$, $\lambda$, and the planetary orbital inclination, $i_p$, we are now able to calculate the system obliquity, $\psi$, from \citep[][]{2005ApJ...631.1215W}
\begin{equation}
\cos \psi = \sin i_{\star} \cos \lambda \sin i_p  + \cos i_{\star} \cos i_p \, .
\end{equation}
In \fref{fig:ob} we show the distribution for $\psi$ when using the obtained distribution for $i_{\star}$ (see \fref{fig:bestfit}), while adopting $i_p$ from \citet[][]{2013ApJ...774L..19V}, and $\lambda$ from \citet{2012ApJ...757...18A}. From the $68\%$ HPD credible region we obtain $83^{\circ}<\psi<111^{\circ}$, consistent with a polar orbit. The corresponding results from using $\lambda$ from \citet{2009ApJ...703L..99W} and \citet{2009PASJ...61L..35N} are given in Table~\ref{tab:orbit}. 

We refer to \citet{2012ApJ...757...18A} for a discussion on the different values for $\lambda$ and the possible reasons for their disagreement \citep[see also][]{2011ApJ...738...50A}. Here we note that regardless which $\lambda$-value we use we find a polar orbit for HAT-P-7 b.

When an assessment of the obliquity is made without knowing the stellar inclination, a flat distribution in $\cos i_{\star}$ is generally assumed for the stellar orientation, as this results in an isotropic distribution for the stellar inclination. From such a distribution it is \emph{a priori} much more likely to observe a random star in an equator-on configuration. In \fref{fig:ob} we show the distribution in $\psi$ from adopting this isotropic distribution in $i_{\star}$. In this way \citet{2009ApJ...703L..99W} estimated $\psi>86.3^{\circ}$ with $99.73\%$ confidence, while \citet{2009PASJ...61L..35N} found $\psi>90^{\circ}$ with $99.70\%$ confidence \citep[both used $i_p$ from][]{2008ApJ...680.1450P}. From this approach a retrograde orbit is thus strongly suggested, but the orbit has a higher probability of being more equatorial than polar. 
With our asteroseismic estimate for the inclination we can substantiate these statistical results as we find $\psi>90^{\circ}$ with $68\%$ credibility using $\lambda$ from \citet{2012ApJ...757...18A}, and now a near polar orbit is the most likely configuration for the system.

\section{Comparison with gyrochronology}
\label{sec:gyro}

Our result for the limits on the stellar rotation rate (see Table~\ref{tab:astro}) can be compared to empirically calibrated gyrochronology relations.
Here we use the form described by \citet[][]{2007ApJ...669.1167B} given as 
\begin{equation}
P(B-V, t) = t^n \times a\left[(B-V)_0 - c \right]^b\, ,
\label{eq:gyro}
\end{equation}
where $t$ is the stellar age in Myr, while $a$, $b$, and $n$ are empirically determined coefficients that vary depending on the calibration set used \citep[see, \eg,][]{2014ApJ...780..159E}.
To compare with this relation, we first need an estimate for the de-reddened colour $(B-V)_0$. The procedure used to get $(B-V)_0$ is described in Appendix~\ref{app:BV}.

\begin{figure}
\includegraphics[width=\columnwidth]{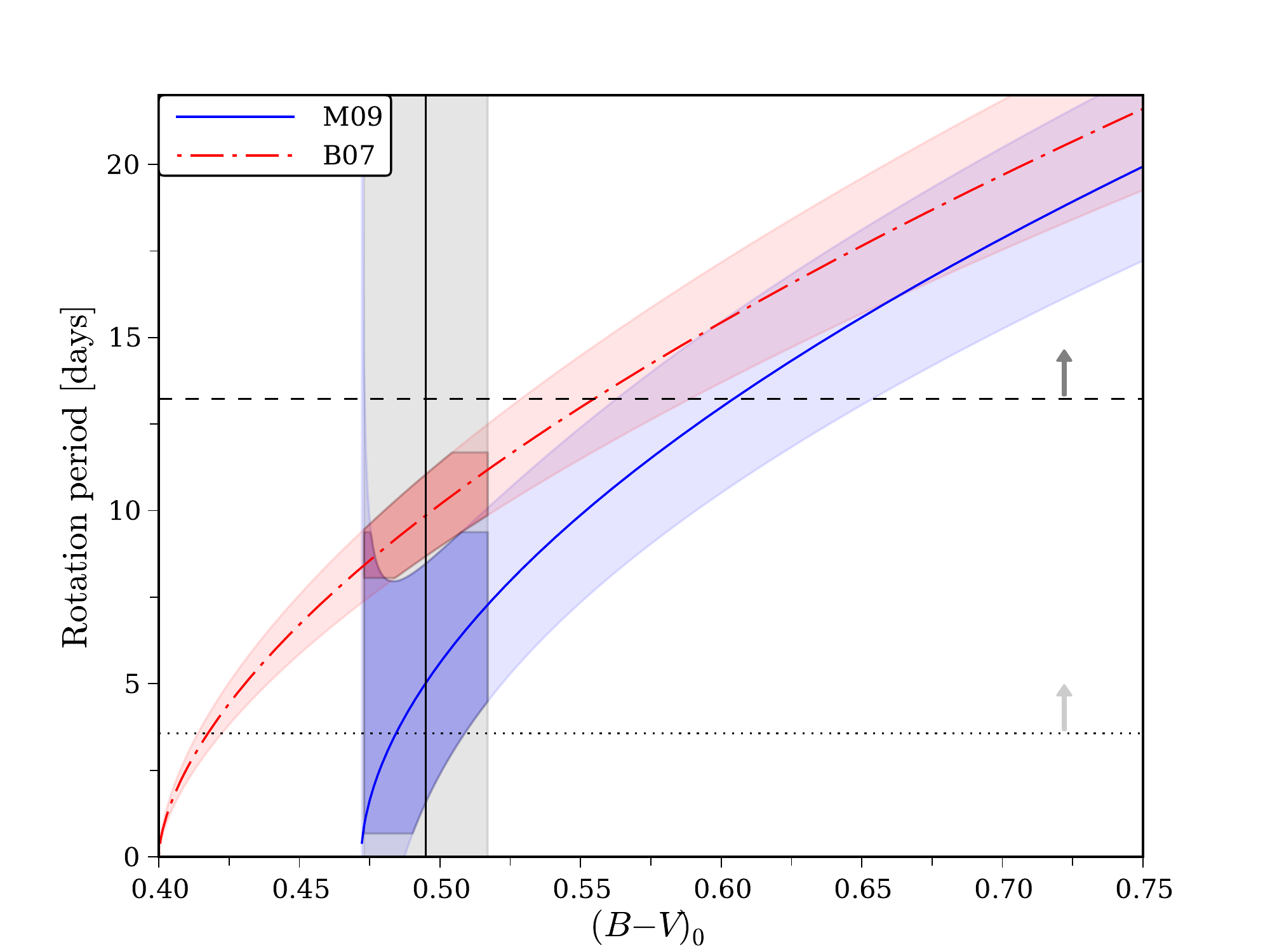}
\caption{\footnotesize Gyrochronology relations from \citet[][]{2009ApJ...695..679M} (M09), \citet[][]{2007ApJ...669.1167B} (B07). Shaded region around each relation (with the same colour) represent the standard error of the relationship between $(B-V)_0$ and rotation period from propagating the uncertainties reported for the coefficients entering \eqref{eq:gyro} together with the uncertainty in our age estimate. The vertical line and shaded region give the value of $(B-V)_0=0.495\pm0.022$ of \hat{}, while darker shaded regions around the relations vertically bound the corresponding uncertainty in the period from the uncertainty in $(B-V)_0$, age, and the gyrochronology relations. Horizontal lines indicates the limits on the $68\%$ (dashed) and $95\%$ (dotted) credible regions on the stellar rotation period from $p$-mode splittings. The upward pointing arrows indicate that these rotation periods are lower limits from the respective HPD credible regions.}
\label{fig:gyro}
\end{figure}

Using the estimate $(B-V)_0=0.495\pm 0.022$ together with the age determined in \sref{sec:modelling} of $t=\rm  2.07\pm 0.36 \, Gyr$ (see Table~\ref{tab:mod_results}; asymmetric uncertainties were added in quadrature), we can estimate the rotation period from the relation in \eqref{eq:gyro}. In \fref{fig:gyro} we show two versions of this relation, namely those by \citet[][]{2007ApJ...669.1167B} (B07), and \citet[][]{2009ApJ...695..679M} \citep[M09; see also][]{2010ApJ...721..675B,2011ApJ...733..115M}. From these we get periods of $9.9\pm 1.8$ (B07) and $5.0\pm 4.0$ days (M09).
The upper value from the B07 relation is comparable to our lower limit period estimate of about $13$ days (see Table~\ref{tab:astro}), so the agreement is not very convincing. On the other hand, the splittings at low inclinations match well this range of rotation periods as the level of ${\sim}4\, \rm \mu Hz$ corresponds to a rotation period of ${\sim}2.9$ days (see \fref{fig:bestfit}). 

We note that a $P_{\rm rot}\propto t^{1/2}$ law might provide a poor description of the rotational evolution for HAT-P-7, as its $(B-V)_0$ puts it in close proximity to the so-called \emph{Kraft break} \citep[][]{1967ApJ...150..551K} where loss of angular momentum via a stellar wind is inhibited by the lack of a sufficiently deep convection zone. If HAT-P-7 is on the low $(B-V)_0$ side of this break the rotation rate becomes a strong function of the initial conditions \citep[see, \eg,][]{2013ApJ...776...67V}, and a gyrochronology scaling is not applicable. 

The close-in hot-jupiter HAT-P-7b has potentially had an impact on the rotation rate of its host star. While a detailed dynamical analysis of this system is beyond the scope of this paper, the synchronisation of rotation and alignment clearly has not been reached yet. The time scale for circularisation has been reached though, with an upper limit on the eccentricity of $e<0.038$ given by \citet[][]{2012MNRAS.422.3151H}. While it is difficult to assess the past interaction between the planet and the star, we note that with the configuration we estimate for the system, \ie, close to a polar orbit of the planet, the rotational synchronisation time will likely be very long as the angular momentum vectors are close to being orthogonal, thereby decreasing the tidal interaction. 
For discussions on the interaction between hot jupiters and their host stars we refer the reader to, \eg, \citet[][]{1981A&A....99..126H}, \citet[][]{2009ApJ...702.1413M}, \citet[][]{2010ApJ...723L..64C}, \citet[][]{2010ApJ...725.1995M}, \citet[][]{2011IAUS..276..238M}, and \citet[][]{2014ApJ...786..102V}.

\section{Activity signatures}
\label{sec:activity}

Activity of stars is linked to the interaction between rotation, convection, and magnetic fields. The interaction can cause magnetic features, such as dark spots and bright faculae, to appear in the photosphere of the stars \citep[see, \eg,][for a review]{lrsp-2005-8}, and plage in the chromosphere. If a star is rotating, such dark or bright regions will induce a temporal modulation of the integrated stellar flux. 

The photospheric features can also cause departures from radiative equilibrium in the stellar chromosphere in plage, and induce emission in the cores of specific spectral lines. For example is emission in cores of Ca \textsc{ii} H\&K lines an often used indicator for magnetic activity \citep[][]{2001A&A...376.1080K,2004ApJS..152..261W,2011A&A...532A..81F,2012A&A...543A.146F}, as it is believed to reflect the amount of non-thermal chromospheric heating above faculae \citep[see, \eg,][for a review]{2008LRSP....5....2H}. For close-in hot-jupiter systems there is furthermore the possibility for a direct magnetic interaction between the planet and the star, where magnetic reconnections, similar to those seen in flares, can cause a heating of the chromosphere \citep[see, \eg,][]{2005ApJ...622.1075S}.

\subsection{Chromospheric activity}

To test for signatures of chromospheric activity we searched the Ca \textsc{ii} H\&K lines in the HIRES spectra from \citet{2009ApJ...703L..99W}, but found no signs of emission or high activity levels. As we see $p$-mode oscillations in HAT-P-7, it must have an outer convection zone, and the $(B-V)_0$ colour derived in \sref{sec:gyro} places it above the limit from \citet[][]{1991ApJ...380..200S} for the onset of activity. The absence of an emission signal could point towards an intrinsically low surface activity caused by a low rotation rate. 

Another effect that would influence the signal is if \hat{} has a low inclination angle. On the Sun, faculae are primarily located in the active latitude bands between around $5$ and $40^{\circ}$ latitude. As faculae and plage are believed to have the same magnetic driver the Ca \textsc{ii} H\&K emission from plage will thus decrease with decreasing inclination from the decreasing projected area of the active regions \citep[][]{2001A&A...376.1080K,2004A&A...413..657F,2007MNRAS.377...17C}.

\subsection{Low frequency region}
\label{sec:peak}

We searched for a signal imparted by a temporal flux modulation at low frequencies in the power spectrum of \hat{} \citep[see, \eg,][]{2011A&A...534A...6C,2013A&A...557L..10N}. Examples of such activity signals in F-type \kp{} stars can for instance be seen in \citet[][]{2014A&A...562A.124M}.

\begin{figure}
\centering
\includegraphics[width=\columnwidth]{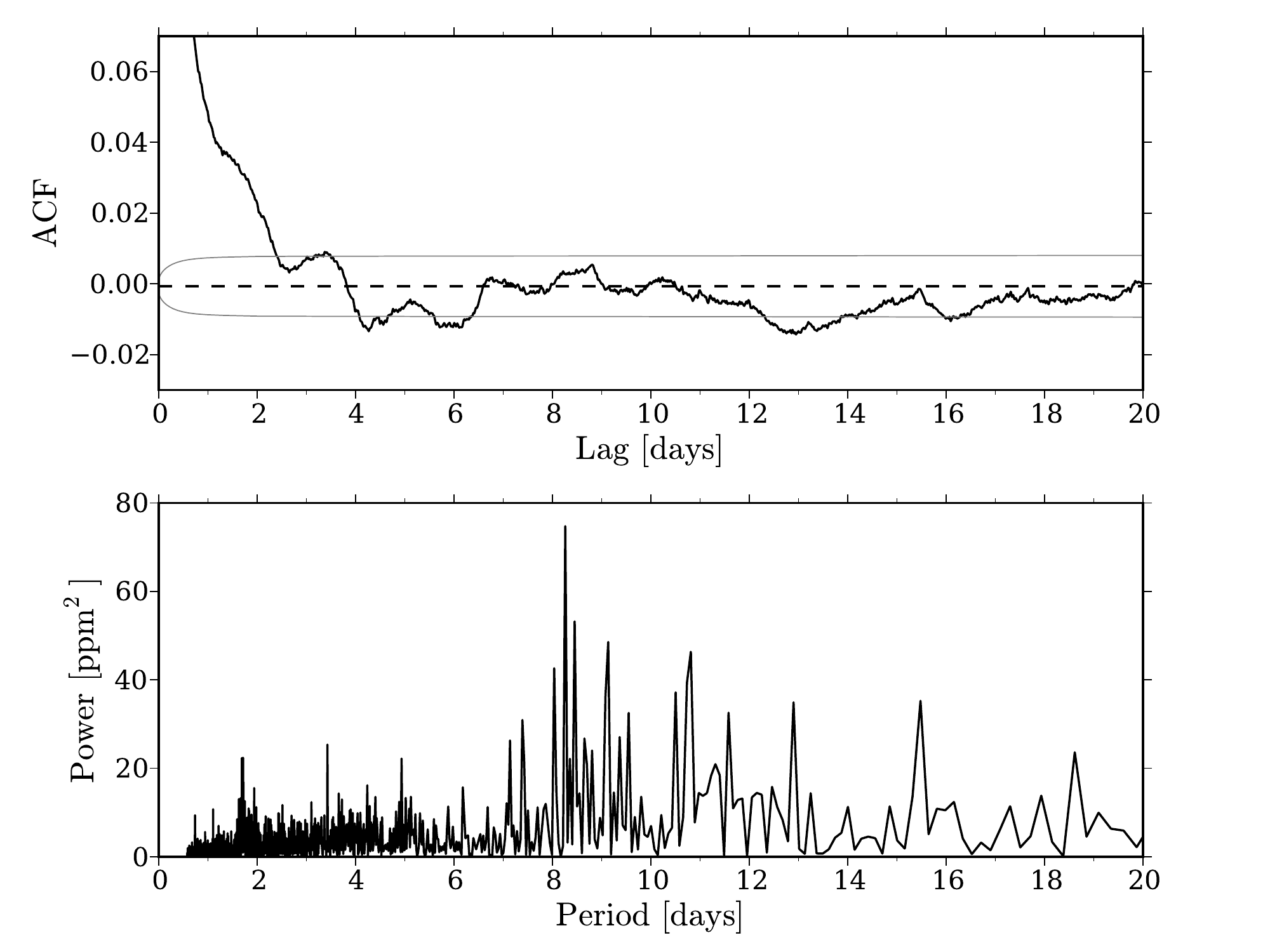}
\caption{\footnotesize Top: autocorrelation function (ACF) of the \hat{}\ time series (black). The dashed horizontal line gives the expectation value for random independent and identically distributed values, while the grey curves give the large-lag $95\%$ confidence levels. Bottom: low-frequency end of the power spectrum in units of period.}
\label{fig:accps}
\end{figure}

The signal from the stellar rotation can also be estimated by, \eg, the autocorrelation function (ACF) of the time series, as shown for instance by \citet[][]{2013MNRAS.432.1203M}.
In \fref{fig:accps} we show the ACF of the corrected time series together with the low-frequency end of the power spectrum (in units of period for convenience). 
The ACF shows no clear sign of modulation. To test the hypothesis that a signal in the time series has died out at lag $k$, we use the large-lag standard error \citep[][]{anderson1976time}, indicated in \fref{fig:accps}. From this it is clear that the signal seen in the ACF is not significant. 
However, the low-amplitude hump at around ${\sim}9$ days in the ACF does seem to align with an increase in power in the power spectrum.

\begin{figure*}
\centering
\includegraphics[width=2\columnwidth]{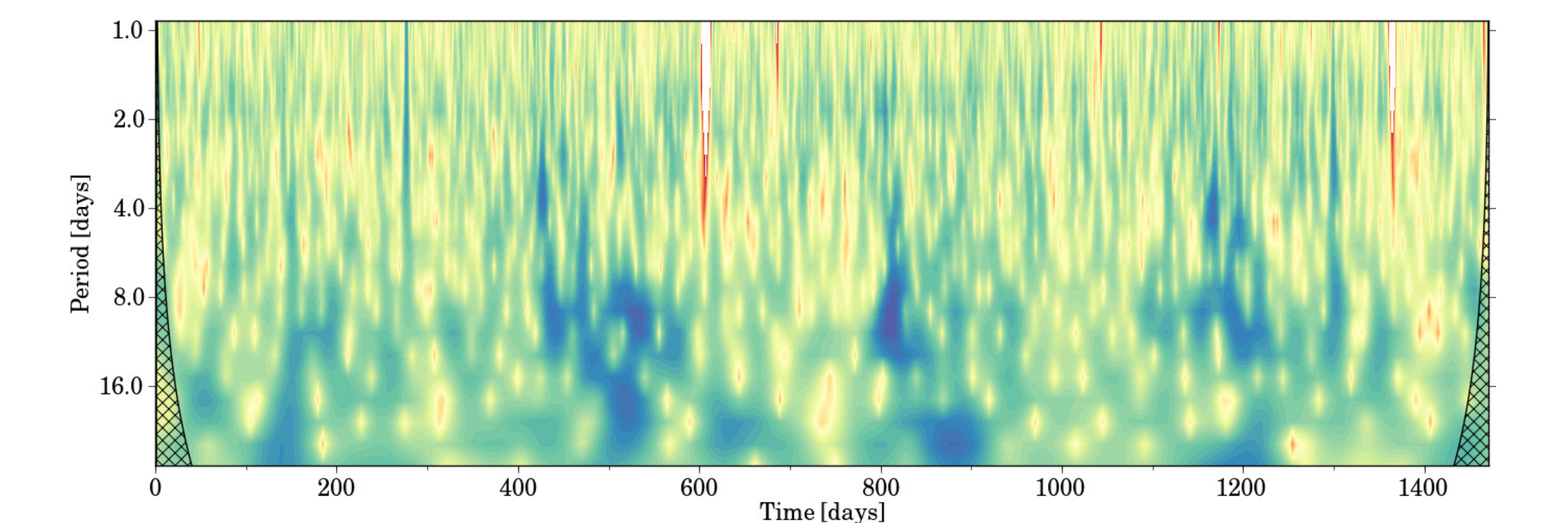}
\caption{\footnotesize Morlet wavelet power spectrum as a function of time for \hat{}, where the time series was binned by 31 points. The colour goes from red at low power to blue at high power - the colours are on a logarithmic scale. The cross-hatched regions at high and low times indicate the \emph{cone of influence}, where edge effects become important \citep[see][]{1998BAMS...79...61T}. No clear signatures from rotation are apparent.}
\label{fig:wave}
\end{figure*}

We investigated the presence of a modulation further via the Morlet wavelet transform \citep[][]{1998BAMS...79...61T} of the time series \citep[see also][]{2013A&A...550A..32M}, this is shown in \fref{fig:wave}. No clear periodicity is seen, and the signal, giving rise to the ${\sim}9$ day hump in the ACF and the power spectrum, seems to be very intermittent. 
Finally we checked the magnetic proxy from \citet[][]{2014ApJ...783..123C}, but also here we did not find indications of a signal.
The fact that a stronger or more localised signal is missing for \hat{} might again be linked to a low stellar inclination. Indeed, if magnetic features such as stellar spots reside primarily near the equator of the star, while we view it close to pole-on, a strong modulation in the light curve would not be expected.

We also note a collection of peaks in the power spectrum around a frequency of $\rm 6.7\pm0.4\, \mu Hz$ (corresponding to a period of ${\sim}41.6$ hours). 
The presence of this power excess is quite robust and is consistently seen when looking at random segments of the total time series and in the wavelet spectrum. We therefore rule out that it originates from random noise. First, we checked that there are no known artefacts at this frequency \citep[][]{kepdatachar}. To see if the signal might originate from another star in the vicinity of \hat{} ($K_p = 10.463$) we located all \kp{} targets with available data within a radius of $3\arcmin$ ($12$ were found). Of these, the star KIC~10666727 ($K_p=13.166$) was found to have strong signatures of spot modulations in its time series, and broadened power excess peaks in its power spectrum around $\rm {\sim}2.2\, \mu Hz$,  $\rm {\sim}4.6\, \mu Hz$, and most interestingly in the region $\rm 6.7-7.2\, \mu Hz$. The angular separation between this target and HAT-P-7 is around $155\arcsec$, which puts the two stars close enough that direct PRF (pixel response function) contamination could occur between them \citep{ref:contamination}. However, according to \citet{ref:contamination}, for stars to contaminate over such a large separation they have to be bright. 
We do not see any significant power excess in HAT-P-7 at other frequencies where KIC~10666727 shows even stronger excess than at $\rm 6.7\pm0.4\, \mu Hz$. Therefore we find it possible, but unlikely, that KIC~10666727 is contaminating the light curve of HAT-P-7. 

\section{Discussion}
\label{sec:dis}

With an asteroseismic modelling we have provided a precise and detailed model for \hat{}, with parameters that can be used in tests of theories on the formation and evolution of planetary systems. These, together with the obliquity, are especially important to constrain when attempting to explain a system such as HAT-P-7 which likely has a close-to polar orbit of its hot-jupiter planet.

Our estimate of the obliquity of the \hat{} system supports the hypothesis described by \citet[][]{2010ApJ...718L.145W} stating that hot-jupiters are born with a large range of obliquities. Planet-planet scatterings \citep[][]{2008ApJ...686..580C}, and the effect of Kozai cycles and tidal friction \citep[][]{2007ApJ...669.1298F} could for example create large initial obliquities. For cool dwarfs with deep convection zones, tidal dissipation would operate efficiently and result in aligned systems. Hot stars fail to align because they lack such a deep convection zone for most of their life on the main-sequence, if not all together. This hypothesis was put forth based on the observed trend of $\lambda$ against $T_{\rm eff}$ where a broad distribution is seen in $\lambda$ for $T_{\rm eff}>6250\, \rm K$, while predominantly low values for $\lambda$ is seen below this temperature. This result was corroborated by \citet{2012ApJ...757...18A} using a larger sample of measurements. These authors also found that the systems with respect to $\lambda$ could be sorted in relative tidal timescales, which in turn depends on, \eg, the planet to star mass ratio, and $a/R_{\star}$ ($a$ being the semimajor axis). While \hat{} falls on the hot side of the dividing temperature between the two regimes, and tidal interactions are expected to be weak, it is worth noting that the temperature is quite close to this dividing line. Also, $a/R_{\star}$ is small which increases tidal interaction \citep{2012ApJ...757...18A}. The relatively slow rotation found suggests that some magnetic braking has and is taking place. The lack of alignment is likely linked to the combined effects of a low mass of the convection zone (making tides ineffective), the close-to orthogonal alignment of the angular momentum vectors, and the relatively young age of the system. Indeed, the star might have started out with a more rapid rotation, where magnetic braking from the developing convection zone has been more effective in slowing down the star than has the tidal interaction in realigning the system. 

With regard to the obliquity of the system, the fact that a third, and possibly a fourth, body is found to be associated \citep{2009ApJ...703L..99W,2012PASJ...64L...7N}, could support the scenario of few-body dynamical interaction early in the life of the system, which resulted in the high obliquity of HAT-P-7b.


\section{Conclusions}
\label{sec:con}

Using asteroseismology we have estimated the stellar inclination and provided new and precise stellar parameters for \hat{} based on the extracted mode frequencies.
Mode frequencies were extracted using a Bayesian MCMC approach to peak-bag the frequency power spectrum from corrected \kp{} data, where we utilised the full SC dataset from Q0-Q17 available from the \kp{} satellite.
From this, information is obtained about stellar, planetary, and system parameters (age, mass, radius, composition, luminosity, mass of convective core) that are important ingredients in, \eg, dynamical simulations used to test theories for the evolution of planetary systems.

We have found the star of the \hat{} system to have a low inclination, $i_{\star}<36.5^{\circ}$ with $68\%$ credibility, meaning that the star is seen close to pole-on. Combining this with estimates for the planetary orbital inclination and the projected obliquity from RM measurements, we find that the close-in hot-jupiter planet is in a high obliquity and likely retrograde orbit (the retrograde solutions account for ${\sim}68\%$ of the PDF in \fref{fig:ob}). Our estimate for the stellar rotation matches empirical findings for stars of the same spectral type, and using the age of the system from our modelling of the oscillation frequencies in combination with an improved estimate for the colour of the star (see Appendix~\ref{app:BV}), yields that estimates from gyrochronology are not in conflict with our results. While the lack of signatures from activity are by no means proof of a low inclination for the star, they are not incompatible with a low inclination. 

To our knowledge this analysis is the first wherein asteroseismology has been able to provide an estimate for $i_{\star}$ that, together with $\lambda$ from RM measurements and $i_p$ from analysis of the transit profile, has allowed for a near complete description of the system geometry. 

For theories attempting to explain the formation and evolution of planetary systems, the \hat{} system is highly interesting, as any such theory must be able to offer an explanation for the system geometry. The analysis presented in this paper has shed more light on the obliquity of the system, and not just the projected obliquity which is normally used in obliquity studies. The result on the obliquity is in agreement with assumptions based on the high value of $\lambda$ and the low value for $v\sin i_{\star}$. Furthermore, the high value for $\psi$ corroborates the theory where the degree of alignment is connected to the tidal evolution of the system \citep[see, \eg,][]{2010ApJ...718L.145W,2012ApJ...757...18A,2014ApJ...786..102V}.

An aspect of the \hat{} system that could be investigated to constrain the obliquity even further is the apparent asymmetry of the transit light curve, seen for instance in the phase curve presented in \citet[][]{2013ApJ...772...51E} and \citet[][]{2013ApJ...774L..19V}. Such an asymmetry is also seen in, \eg, the KOI-13 system, and it was found by \citet[][]{2011ApJ...736L...4S} and \citet[][]{2011ApJS..197...10B} to be in good agreement with the predictions by \citet[][]{2009ApJ...705..683B} for a planet crossing over the gravity brightened polar region of its rapidly rotating host star on a high obliquity orbit.


\begin{acknowledgements} 
The authors wish to thank the entire \kp{} team, without whom these results would not be possible.
We are thankful to O. Benomar and his collaborators for their approach to the contemporaneousness of our respective studies.
We would like to thank Eric Stempels and Heidi Korhonen for useful discussions in the early phases of the project.
Thanks to Joshua N. Winn and John A. Johnson for making their Keck spectra available to us.
Funding for the Stellar Astrophysics Centre (SAC) is provided by The Danish National Research Foundation. The research is supported by the ASTERISK project (ASTERoseismic Investigations with SONG and \kp{}) funded by the European Research Council (Grant agreement no.: 267864).
M.L. would like to thank the asteroseismology group at SIfA for their hospitality during a research stay where some of this work was carried out. 
M.L. also wishes to thank Niels Bohr Fondet for financial support for the research stay at SIfA.
T.L.C., G.R.D., W.J.C., and R.H. acknowledge the support of the UK Science and Technology Facilities Council (STFC).
C.K. acknowledge the support of the Villum Foundation.
M.B.N. acknowledges research funding by Deutsche Forschungsgemeinschaft (DFG) under grant SFB 963/1 ``Astrophysical flow instabilities and turbulence'' (Project A18).
The research leading to these results has received funding from the European Community's Seventh Framework Programme (FP7/2007-2013) under grant agreement no. 269194.
This research has made use of the following web resources: the SIMBAD database (simbad.u-strasbg.fr), operated at CDS, Strasbourg, France; NASAs Astrophysics Data System Bibliographic Services (adswww.harvard.edu); arxiv.org, maintained and operated by the Cornell University Library.
\end{acknowledgements}


\bibliography{MasterBIB,mia,victor}


\onecolumn
\appendix

\section{Power spectrum modelling and optimisation}
\label{app:modelling_power}


\subsection{Modelling the power spectrum}

To describe the observed power spectral density of a mode peak in the frequency power spectrum, we use a standard Lorentzian function \citep[see, e.g.,][]{1990ApJ...364..699A, 2003ApJ...589.1009G} given by
\begin{equation}
L_{nl m}(\nu) = H_{nl m} \left[1+\left(\frac{(\nu-\nu_{nl m})}{\Gamma_{nl}/2}\right)^2 \right]^{-1}.
\label{eq:lorentz}
\end{equation}
The use of a Lorentzian function for the line profile comes from the nature of solar-like $p$-modes, namely that they are stochastically driven by turbulent convection in the outer envelope after which they are intrinsically damped \citep[see, \eg,][]{1994ApJ...424..466G}. 
In this equation $H_{nl m}$ is the mode height, $\nu_{nl m}$ is the resonance frequency of the mode, while $\Gamma_{nl}$ is a measure of the damping rate of the mode and gives the FWHM of $L_{nl m}(\nu)$. 

In the case of slow stellar rotation the star is generally assumed to rotate as a rigid body and the modes will to first order be split as \citep{1951ApJ...114..373L}
\begin{equation}
\nu_{nl m} = \nu_{nl} + m\frac{\Omega}{2\pi}(1-C_{nl}) \approx \nu_{nl} + m\nu_s.
\label{eq:split}
\end{equation}
Here $m$ is the azimuthal order of the mode, $\Omega$ is the angular rotation rate of the star, and $C_{nl}$ is a dimensionless constant that describes the effect of the Coriolis force (the Ledoux constant). For high-order low-degree solar-like oscillations, like the ones we want to analyse, this quantity is of the order $C_{nl} < 10^{-2}$, and is therefore neglected. In this way we see that the splitting due to rotation between adjacent components of a multiplet will to a good approximation be given by $\nu_s = \Omega/2\pi$.

In assuming equipartition of power between the components of a multiplet (\ie\ no assumed preference in the excitation for prograde over retrograde propagating modes), it is possible to calculate the geometrical modulation of the relative visibility between the $2l+1$ multiplet components as a function of $i_{\star}$ as \citep[see, \eg,][]{1977AcA....27..203D, 2003ApJ...589.1009G}
\begin{equation}
\mathcal{E}_{l m}(i_{\star}) = \frac{(l - |m|)!}{(l + |m|)!}\left[P^{|m|}_{l} (\cos i_{\star}) \right]^2\, ,
\label{eq:epsilon}
\end{equation}
where $P^{m}_{l}(x)$ are the associated Legendre functions.

With this, the limit spectrum (noise free) to be fit to the power spectrum is as expressed in \eqref{eq:limitspec} (see \fref{fig:Powerspecs_full}) . 
Comparing Eqs.~\ref{eq:lorentz} and \ref{eq:limitspec} it is seen that the height is given by $\mathcal{E}_{l m}(i)S_{nl}$. By assuming equipartition of energy between different radial orders this can be written as
\begin{equation}
H_{nl m} = \mathcal{E}_{l m}(i_{\star})S_{nl} = \mathcal{E}_{l m}(i_{\star}) \tilde{V}_{l}^2 \alpha_{l=0}(\nu)\, .
\label{eq:vis}
\end{equation}
The factor $\tilde{V}_{l}^2$ is a measure of the relative visibility in power (primarily set by partial cancellation) between non-radial and radial ($l=0$) modes, while $\alpha_{l=0}(\nu)$ represents a (mainly) frequency dependent height modulation for the radial modes, generally represented by a Gaussian centred on the frequency of maximum oscillation power, \numax. 


\subsection{Optimisation procedure}
\label{sec:Optimisation_procedure}

The fitting of \eqref{eq:limitspec} to the power spectrum is done in a Bayesian manner by mapping the \emph{posterior probability}:
\begin{equation}
p(\mathbf{\Theta}| D,I) \propto {p(\mathbf{\Theta}|I)p(D|\mathbf{\Theta},I)} \, . 
\end{equation}
Here $p(\mathbf{\Theta}|I)$ is the \emph{prior probability} assigned to the parameters $\mathbf{\Theta}$ from any prior information \emph{I}, and $p(D|\mathbf{\Theta},I)$ is the \emph{likelihood} of the observed data \emph{D} given the parameters $\mathbf{\Theta}$.
The posterior is approximated using the affine invariant MCMC sampler \texttt{emcee} (\citet[][]{2013PASP..125..306F}, see also \citet{2012ApJ...745..198H}). Using the \texttt{emcee} routine, the posterior distribution is mapped after which parameter estimates are evaluated as the median of the respective marginalized distributions (see \sref{sec:res}). In the sampling we enable the parallel tempering scheme of \texttt{emcee} and use five temperatures with \emph{tempering parameters} set as $\beta_i = 1.2^{1-i}$ \citep[][]{2009A&A...506...15B,2011A&A...527A..56H}. The affine invariant character of the \texttt{emcee} sampler ensures that it works efficiently in spite of linear parameter correlations which are a problem for many MCMC algorithms. 
In our optimisation we make both a fit to a large (full fit) and a small (small fit) frequency range (see \sref{sec:res} for details). We employ $1500$ (full fit) and $2000$ (small fit) \emph{walkers}, respectively, all initiated from a sampling of the prior distributions. Each walker is stopped after $10\,000$ steps after which we thin the chains by a factor of 10 (full fit) or 5 (small fit). We cut away a burn-in part of each chain based on the Geweke\footnote{Using the \texttt{geweke} module of PyMC \citep[][]{ref:PyMC}.} statistics \citep[][]{MR1380276}, and check for good mixing using the autocorrelation time of the chains and by performing a visual inspection of the \emph{traces} of walkers in parameter space. We refer the reader to \citet[][]{2011A&A...527A..56H} and references therein for further details on the MCMC nomenclature and \citet[][]{2013PASP..125..306F} for the specifics of the \texttt{emcee} sampler. 
To ensure better numerical stability we map the logarithm of the posterior with the description of the log-likelihood function from \citet[][]{1990ApJ...364..699A} and \citet{1994A&A...289..649T}.
With regard to priors, we use \emph{top-hat} priors for location parameters (\eg, $\nu_{nl}$) and scale invariant \emph{modified Jeffryes'} priors for scale parameters (\eg, $S_{n0}$). 
To decrease the computation time the limit-spectrum (\eqref{eq:limitspec}) was only fit to the frequency-range including the identified oscillation modes (see \sref{sec:res_overall} for further details).
To better constrain the stellar noise-background in the relatively narrow range occupied by the oscillation modes, \eqref{eq:bg} was first fit to the power spectrum in the frequency range from $\rm 100-8496\, \mu Hz$ (the upper limit is the approximate Nyquist frequency of SC data) and included either one or two characteristic time scales corresponding to the contributions from granulation only or granulation and faculae (this lower-limit frequency ensures that the activity component can be omitted). We also added a Gaussian function to \eqref{eq:bg} to account for the power excess from solar-like oscillations seen in \hat{}. Using the \emph{deviance information criterion} \citep[DIC;][]{MR1979380} we found that only one component is needed in the description of the background. The medians of the posteriors from this fit were then used to fix the background in the fit of \eqref{eq:limitspec}. 


\section{Linewidths and visibilities}
\label{app:lwvis}

\subsection{Linewidths}
\begin{figure}[h]
\centering
\includegraphics[width=0.5\columnwidth]{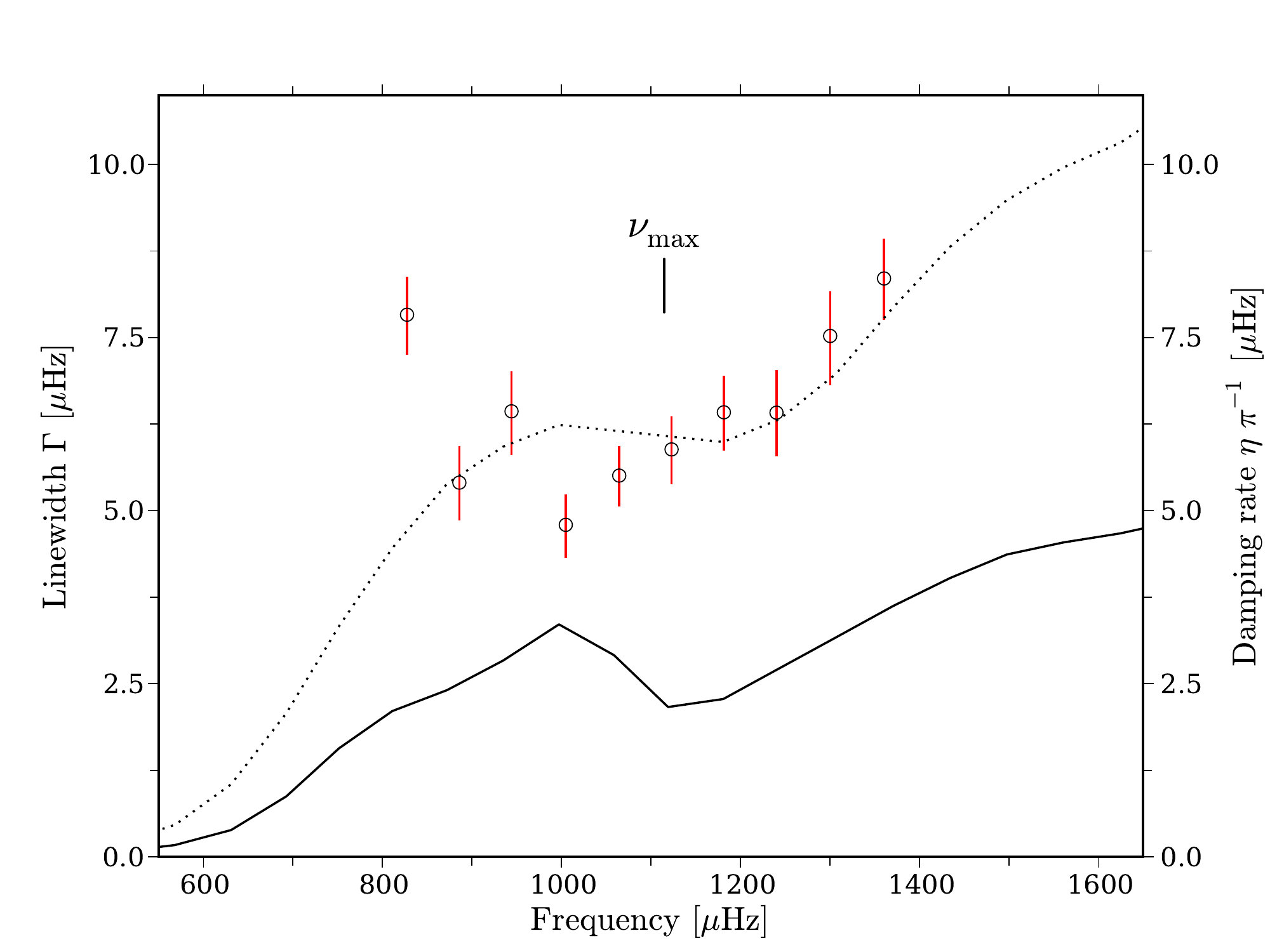}
\caption{\footnotesize Measured linewidths for radial order modes are given by the open symbols as a function of frequency (left axis), and plotted with associated errors. 
Theoretical linear damping rates, multiplied by two, are shown by the solid curve (right axis). The dotted curve shows smoothed damping rates, multiplied by a factor of $2.2$.
Inputs to the theoretical model were taken from the best-fit GARSTEC model, \eg\ the radius at the base of the surface convection zone ($R_{\rm bcz}/R_{\star} = 0.862$). }
\label{fig:linewidth}
\end{figure}

The linewidths can have a large impact on the estimated splitting as a small splitting could be equally well fit by a slightly larger linewidth. This is especially a problem when the splitting is smaller than the linewidth.
We show in \fref{fig:linewidth} our fitted linewidths for the radial ($l=0$) modes and their associated uncertainties. 
As a sanity check, we may first compare the mode linewidth at the frequency of maximum power, $\nu_{\rm max}$, to the estimate from Eq.~(2) of \citet[][]{2012A&A...537A.134A} and using combined values from their Table~$2$. 
Using the spectroscopic temperature of $T_{\rm eff}= 6350\pm126\rm \, K$, one gets $\rm \Gamma \approx 3.5 \pm 1.1\,\mu Hz$. From the three $l=0$ modes closest to $\nu_{\rm max}$ we find (central) values of $\rm \Gamma$ between $5.5$ and $6.4\rm\,\mu Hz$ (see \fref{fig:linewidth}), thus higher than expected from the \citet[][]{2012A&A...537A.134A} formulae. 
%

Additionally, we estimated linear damping rates, $\eta$, which we assume to be approximately equal to half of the observed linewidths, 
\ie\ $\Gamma\simeq\eta\,\pi^{-1}$, if $\eta$ is in units of angular frequency. The outcome is shown in \fref{fig:linewidth}. The computations included a full nonadiabatic treatment of the pulsations and convection dynamics. Both the convective heat and momentum (turbulent pressure) fluxes were 
treated consistently in both the equilibrium and pulsation computations using the nonlocal generalization of the time-dependent 
convection model by \citet[][]{1977ApJ...214..196G} and \citet[][]{1977LNP....71...57G}. The computations were carried out as described by \citet[][]{1999A&A...351..582H} and \citet[][]{2002MNRAS.336L..65H}. For the nonlocal convection parameters we adopted the values $a^2=900$ and $b^2=2000$ and the mixing-length paramter was calibrated to obtain the same depth of the surface convection zone as in the best GARSTEC model. For the anisotropy parameter, $\Phi$ \citep[see][]{2002MNRAS.336L..65H},  
the value $2.50$ was adopted. 

After applying a median smoothing filter to $\eta\,\pi^{-1}$, with a width in frequency corresponding to five radial modes, the result of $\rm 2.8\,\mu Hz$ at $\nu_{\rm max}$ of the oscillation heights lies within the error bars of the observational scaling relation by \citet[][]{2012A&A...537A.134A}. In order to fit the observations in Figure B.1 we multiplied the median-smoothed estimates by a factor of 2.2 (dotted curve in \fref{fig:linewidth}), a factor that is in agreement with previous comparisons between linewidth observations and model estimates for hotter solar-like stars \citep[see, \eg,][]{2006ESASP.624E..28H,2012ASPC..462....7H}.

\subsection{Visibilities}
For the visibilities (see Appendix~\ref{app:modelling_power}), we estimate $\tilde{V}^2_1=1.39\pm 0.08$ and $\tilde{V}^2_2=0.46 \pm 0.07$ from the small fit. These are in reasonable agreement with the theoretical values of $\tilde{V}^2_1\approx 1.51$ and $\tilde{V}^2_2~\approx~0.53$ estimated from the tables of \citet[][]{2011A&A...531A.124B}. We do note, however, that such an agreement is no guarantee for correct values as some stars do show deviations from the simple theoretical estimates \citep[see, \eg,][]{2014ApJ...782....2L}.
If we calculate the visibilities using the method described in \citet[][]{2011A&A...531A.124B} but adopt a quadratic limb-darkening (LD) law and measured LD-parameters from fits to the planetary transit by \citet[][]{2013ApJ...774L..19V} and \citet[][]{2013ApJ...764L..22M}, we obtain values that are slightly lower than the ones from theoretical LD parameters: $\tilde{V}^2_1\approx1.46\pm 0.02$ and $\tilde{V}^2_2~\approx~0.46\pm0.01$. These values agree within the errors with the fit values, which is encouraging given the very simplified assumptions adopted in the \citet[][]{2011A&A...531A.124B} calculation, where for instance all non-adiabatic effects are neglected.


\section{Determining $(B-V)_0$}
\label{app:BV}

We determined the photometric stellar parameters for \hat{} by combining asteroseismic results with the InfraRed Flux Method (IRFM) \citep[see][]{2011ApJ...740L...2S,2012ApJ...757...99S}. We adopted our seismic $\log g$ and the spectroscopic metallicity from \citet{2013ApJ...767..127H,2014ApJS..211....2H} and used the IRFM implementation described in \citet[][]{2014ApJ...787..110C}, where different three-dimensional reddening maps are used to constrain extinction. At a distance of $320$ pc \citep[approximate distance to HAT-P-7 determined by][]{2008ApJ...680.1450P}, reddening varies between $0.02 < \mathrm{E} \left( \mathrm{B} - \mathrm{V} \right) < 0.03$.

Unfortunately, optical measurements of \hat{} are quite uncertain, and depending whether the Tycho2 \citep[][]{2000A&A...355L..27H} or APASS \citep[][]{2009AAS...21440702H} photometry is used, the resulting $T_{\rm eff}$ will vary anywhere between $6350$ and $6650$ K. At the magnitude of our star, \mbox{Tycho2} photometry starts to get increasingly uncertain \citep[][]{2000A&A...355L..27H} (although its $T_{\rm eff}$ would be in overall good agreement with the spectroscopic estimate of $T_{\rm eff} = 6350 \pm 126$ K), while the APASS $(B-V)_0$ index (\ie, after correcting it for reddening) is almost as red as the solar one \citep[][]{2012ApJ...752....5R}, thus suggesting a $T_{\rm eff}$ close to solar. The higher $T_{\rm eff}$ is, however, confirmed by the (reddening corrected) $J-K_s$ index of HAT-P-7 \citep[indeed bluer than the solar one;][]{2012ApJ...761...16C}. From these considerations we thus discard the APASS $(B-V)_0$ as faulty, and adopt a photometric $T_{\rm eff} = 6500 \pm 150$ K, where the generous errors account for the discussed uncertainties. With this $T_{\rm eff}$ and $\rm [Fe/H]=0.26$, we can invert the colour-$T_{\rm eff}$-$\rm [Fe/H]$ relation of \citet[][]{2010A&A...512A..54C}, which returns an intrinsic (\ie, unreddened) colour of $(B-V)_0=0.455 \pm 0.040$ mag. Should the spectroscopic $T_{\rm eff}$ and uncertainty be adopted, then its $(B-V)_0=0.495 \pm 0.022$.

Synthetic photometry offers an alternative way of assessing the $(B-V)_0$ colour for \hat{} (all synthetic quantities are obtained by interpolating at the "known" physical parameters of the star, and thus unaffected by reddening). We use the large grid of MARCS \citep[][]{2008A&A...486..951G} synthetic colours and interpolation routines provided by \citet[][]{2014arXiv1407.6095C} to infer the $(B-V)_0$ index of \hat{}, should the spectroscopic parameters and asteroseismic $\log g$ be adopted. 

To estimate the uncertainty in the synthetic $(B-V)_0$, we also compute all possible $T_{\rm eff}$ and $\rm [Fe/H]$ combinations allowed by the spectroscopic uncertainties (while the seismic $\log g$ is so precisely known that changing it makes no difference). With this procedure, we obtain $(B-V)_0 = 0.498 \pm 0.020$, in excellent agreement with the estimate from the empirical colour-$T_{\rm eff}$-$\rm [Fe/H]$ relation when using the spectroscopic $T_{\rm eff}$.

For the gyrochronology calculation we will use the value $(B-V)_0=0.495 \pm 0.022$ from the IRFM when using the spectroscopic $T_{\rm eff}$.

\newpage
\section{Peak-bagging results}

\begin{table}[bh]
\begin{center}
\begin{threeparttable}
\caption{Frequencies extracted from the MCMC peak-bagging.}
\begin{tabular}{llll}  
\toprule \\ [-0.25cm] 
$n$ & $l=0$ $\rm [\mu Hz]$ & $l=1$ $\rm [\mu Hz]$ & $l=2$ $\rm [\mu Hz]$ \\[0.15cm] 
\midrule
$ 9$  &								& $623.18^{+0.35}_{-0.43}$ &  \\[0.15cm]
$ 10$ & $651.42^{+0.50}_{-0.48}$  	& $683.36^{+0.38}_{-0.41}$ & $710.83^{+0.56}_{-0.55}$ \\[0.15cm]
$ 11$ & $715.37^{+0.34}_{-0.34}$ 	& $740.71^{+0.35}_{-0.35}$ & $767.85^{+1.01}_{-0.95}$ \\[0.15cm]
$ 12$ & $771.46^{+0.79}_{-0.87}$ 	& $796.46^{+0.42}_{-0.41}$ & $824.85^{+1.04}_{-0.91}$ \\[0.15cm]
$ 13$ & $827.89^{+0.45}_{-0.51}$ 	& $853.91^{+0.27}_{-0.26}$ & $882.37^{+0.69}_{-0.61}$ \\[0.15cm]
$ 14$ & $886.13^{+0.28}_{-0.29}$ 	& $911.71^{+0.23}_{-0.23}$ & $940.68^{+0.66}_{-0.59}$ \\[0.15cm]
$ 15$ & $944.26^{+0.39}_{-0.43}$ 	& $971.74^{+0.18}_{-0.19}$ & $1000.49^{+0.55}_{-0.57}$ \\[0.15cm]
$ 16$ & $1005.00^{+0.24}_{-0.25}$ 	& $1031.51^{+0.17}_{-0.16}$ & $1059.36^{+0.51}_{-0.54}$ \\[0.15cm]
$ 17$ & $1064.86^{+0.25}_{-0.27}$ 	& $1090.92^{+0.17}_{-0.17}$ & $1118.71^{+0.41}_{-0.41}$ \\[0.15cm]
$ 18$ & $1123.36^{+0.27}_{-0.27}$ 	& $1149.77^{+0.19}_{-0.18}$ & $1177.55^{+0.71}_{-0.75}$ \\[0.15cm]
$ 19$ & $1181.74^{+0.29}_{-0.28}$ 	& $1208.19^{+0.20}_{-0.20}$ & $1236.32^{+0.58}_{-0.53}$ \\[0.15cm]
$ 20$ & $1240.64^{+0.32}_{-0.31}$ 	& $1267.67^{+0.23}_{-0.25}$ & $1297.08^{+0.89}_{-0.73}$ \\[0.15cm]
$ 21$ & $1300.45^{+0.47}_{-0.53}$ 	& $1327.44^{+0.30}_{-0.31}$ & $1356.65^{+0.74}_{-0.64}$ \\[0.15cm]
$ 22$ & $1360.68^{+0.50}_{-0.53}$ 	& $1388.21^{+0.38}_{-0.37}$ & $1416.94^{+1.02}_{-0.96}$ \\[0.15cm]
$ 23$ & $1421.92^{+0.76}_{-0.77}$ 	& $1448.75^{+0.50}_{-0.47}$ & $1478.18^{+1.21}_{-1.21}$ \\[0.15cm]
$ 24$ & $1482.68^{+0.90}_{-0.92}$ 	& $1510.15^{+0.55}_{-0.58}$ & $1539.50^{+0.93}_{-1.15}$ \\[0.15cm]
$ 25$ & $1545.14^{+0.84}_{-0.94}$ 	& $1568.95^{+0.64}_{-0.64}$ & $1595.15^{+0.77}_{-1.30}$ \\[0.15cm]
$ 26$ & $1602.95^{+0.46}_{-1.23}$ 	&  \\[0.15cm]
\bottomrule
\end{tabular}
\label{tab:mcmc}
\begin{tablenotes}
	\tiny
	\item Note: From the \'{e}chelle diagram in \fref{fig:model_compare} it seems that the signal fitted as the lowest $l=0$ mode might originate from an $l=2$ mode. 
\end{tablenotes}
\end{threeparttable}
\end{center}
\end{table}

\end{document}